\documentclass[aps,groupaddress,nofootinbib]{revtex4}

\usepackage{slashed}
\usepackage{braket}
\usepackage{graphicx}
\usepackage{amsmath}
\usepackage{amssymb}
\usepackage{epsfig}
\usepackage{hhline}
\usepackage{soul}
\usepackage{tabularx,pifont,multirow}
\usepackage{enumitem}
\usepackage{colordvi}
\usepackage{xcolor}

\newcommand{\be}{\begin{equation}}
\newcommand{\ee}{\end{equation}}
\newcommand{\bea}{\begin{eqnarray}}
\newcommand{\eea}{\end{eqnarray}}

\newcommand{\nicktext}[1]{{\textcolor{black}{#1}}}

\newcommand{\tmop}[1]{\ensuremath{\operatorname{#1}}}

\begin{document}
% Input the symbols file
% Preprints
\preprint[\leftline{KCL-PH-TH/2020-{\bf 34}}

% This is needed to format the full author list
%\long\def\inst#1{\par\nobreak\kern 4pt\nobreak
%{\it #1}\par\vskip 10pt plus 3pt minus 3pt}
%

% Title of the paper
\title{ \bf
\boldmath
\Large     Dynamical Majorana Neutrino Masses and Axions II: \\ Inclusion of Axial 
Background and Anomaly Terms} % end title{}

% Input author list file

\author{Nick E.~Mavromatos} 

\vspace{0.1cm}
\affiliation{Theoretical Particle Physics and Cosmology Group, Department of Physics, King's College London, Strand, London WC2R 2LS, UK}

\author{Alex Soto}
 
\affiliation{Instituto de Fisica, Pontificia Universidad Catolica de Chile
\\Vicu$\tilde n$a Mackenna 4860, Santiago 7820436, Chile, and \\ Theoretical Particle Physics and Cosmology Group, Department of Physics, King's College London, Strand, London WC2R 2LS, UK
}

%\date{\today}

\bigskip 

\begin{abstract}

\begin{center}
\textbf{Abstract }
\end{center}

We extend the study of a previous publication~\cite{amsot} on Schwinger-Dyson dynamical mass generation for fermions and pseudoscalar fields (axion-like particles (ALP)), in field theories containing Yukawa type interactions between the fermions and ALPs, by incorporating anomaly terms and/or (constant) axial background fields. The latter are linked to some Lorentz (and CPT) violating scenarios for leptogenesis in the early Universe. We discuss both Hermitian and non-Hermitian Yukawa interactions and axial backgrounds, which are motivated in the context of some scenarios for radiative (anomalous) Majorana sterile neutrino masses in some effective field theories, including attractive four-fermion interactions. The reality requirement for the radiative, anomalously generated, mass component for the fermions, restricts our considerations to the cases where Yukawa interactions and the anomaly terms are either  both hermitian or both  anti-hermitian. 
We show that, for a Hermitian Yukawa interaction,  there is no (pseudo)scalar dynamical mass generation, but there is fermion dynamical mass generation, provided one adds a bare (pseudo)scalar mass. For this case, the hermitian anomaly terms play a similar r\^ole in inducing dynamical mass generation for fermions as the four-fermion attractive interactions, and as such they can themselves generate a small dynamical mass. For antihermitian Yukawa interactions, an antihermitian anomaly resists mass generation. The axial background terms assist  dynamical mass generation induced by antihermitian Yukawa interactions, in the sense that the larger the magnitude of the background, the larger the dynamical mass. For hermitian Yukawa interactions, however, the situation is the opposite, in the sense that the larger the background the smaller the dynamical mass. 
We also compare the anomaly-induced dynamical mass with the radiative fermion mass in models of sterile neutrinos, and find that in  cases where the dynamical mass occurs, the latter dominates over the anomalously generated radiative sterile-neutrino mass.

\end{abstract}
\maketitle
%\newpage

\section{Introduction and Motivation \label{sec:intro}}

In a previous article~\cite{amsot}, we have discussed dynamical mass generation in field theory models with weak Yukawa interactions, in the presence of attractive four fermion contact interactions. 
The work was motivated by some string-inspired models involving anomalously generated Majorana fermion masses~\cite{mp}, where the small Yukawa couplings of pseudoscalar axions with Majorana neutrinos are assumed to be induced by shift-symmetry breaking non-perturbative effects, such as string instantons.

The effective action, describing the interaction of (canonically normalised) axion and sterile neutrino fields reads~\cite{mp} 
\begin{eqnarray} \label{bacoupl3} 
\mathcal{S}_a \!&=&\! \int d^4 x
    \sqrt{-g} \, \Big[\frac{1}{2} (\partial_\mu a(x) )^2  - \frac{\gamma\, c_1\,
        a(x)}{192 \pi^2 f_b \sqrt{1 - \gamma^2}}
      {R}^{\mu\nu\rho\sigma} \tilde{R}_{\mu\nu\rho\sigma} \nonumber\\ 
&&\hspace{-5mm} + \frac{i\, \lambda}{\sqrt{1 - \gamma^2}} \, 
a(x)\, \Big( \overline{\psi}_R^{\ C} \psi_R - \overline{\psi}_R
\psi_R^{\ C}\Big)  
    + i\,  \sum_{j} \, \overline \psi_j \slashed{\nabla} \, \psi_j   + \dots \Big]\;,
\end{eqnarray}
where $a(x)$ denotes an axion field (which in the context of the string-effective model of \cite{mp} 
could be different from the QCD axion, {\it e.g.}, associated with string moduli fields~\cite{arvanitaki}, 
$\psi_R^{\ C} = (\psi_R)^C$ is the (Dirac) charge-conjugate of the right-handed fermion $\psi_R$,
which in \cite{mp} is considered to be a sterile neutrino, 
$\nabla_\mu $ denotes the (torsion-free) gravitational covariant derivative, and the index $j$ in $\psi_j$ runs over fermion species $\psi_j$, including the right-handed (sterile)  fermions $\psi_R$. 
The quantity 
\begin{align}\label{fb}
f_b \equiv (3\kappa^2/8)^{-1/2} = \frac{M_{\rm Pl}}{\sqrt{3\pi}}
\end{align}
with $M_{\rm Pl}$ the reduced Planck constant, 
plays the r\^ole of the axion constant in this model.
The $\dots $ in \eqref{bacoupl3} denote 
{\it repulsive} four-fermion self-interaction terms  involving the (square of the) axial fermion current, $J^{5\,\mu} \,  \equiv \sum_j \overline \psi_j \gamma^\mu \gamma^5 \psi_j$, where the sum is over all fermion species, including the right-handed fermions. Such repulsive four fermion  terms arise after integrating out the (quantum) torsion fields of the effective field theory provided by the field strength of the antisymmetric spin-1 tensor field that exist in the massless gravitational string multiplet of the underlying string theory~\cite{mp}. They will not be of relevance to us in this work. 

\nicktext{In \cite{amsot} we also explained the reasons why we ignore axion potentials in this analysis, except perhaps bare axion mass terms. Here 
we deal with stringy and not ordinary axions, stemming from appropriate flux forms of the underlying string theory~\cite{arvanitaki,mp}. Due to underlying gauge stringy symmetries, such stringy axions are not characterised by any potentials in string perturbation theory. Non-perturbative (e.g. string instanton) effects can generate such potentials. However, we 
assume that our Yukawa-interaction induced dynamical masses occur at epochs of the Universe that precede the generation of such axion potentials through non perturbative stringy effects, except perhaps bare axion mass terms, which as we explained in \cite{amsot} could be generated by physics deep in the quantum regime of string theory. Such physics might be the same as the one that generates the shift-symmetry-breaking Yukawa interactions.} 

The coefficient $c_1$ in the gravitational anomaly~\cite{anomalies} term depends on the number of chiral fermions that circulate in the anomaly loop.  We remind the reader that, in terms of the anomalous axial current for massless chiral fermions, the gravitational anomaly terms 
are given by~\cite{anomalies} 
\be\label{anomfermion}
\int d^4x \sqrt{-g}\,  \frac{\gamma\, c_1\,
        a(x)}{192 \pi^2 f_b \sqrt{1 - \gamma^2}}\, {R}^{\mu\nu\rho\sigma} \tilde{R}_{\mu\nu\rho\sigma} = -\, \int d^4x \sqrt{-g}\, 
\frac{\gamma\, a(x)}{f_b \sqrt{1 - \gamma^2}} \, \nabla_\mu J^{5\mu}  = \int d^4x \sqrt{-g}\, \frac{\gamma\,\partial_\mu a(x)}{f_b \sqrt{1 - \gamma^2}} \,  J^{5\mu}\,,
 \ee
where the last equality is obtained by partial intergration, assuming that the fields and their derivatives vanish at infinity. For the purposes of our work we ignore gauge anomalies, as we restrict ourselves to sterile right-handed fermions (neutrinos), which in our concrete examples discussed here and in \cite{amsot}, are assumed to be the only chiral fermion species circulating in the anomaly loop diagram. 
In this case one may set $c_1=1$.
 
The Yukawa coupling $\lambda $ in \eqref{bacoupl3} is assumed to be due to non perturbative string instanton effects, and as such is expected to be  small. These
terms \emph{break explicitly} the shift symmetry $a(x) \to a(x) + c$, in the same spirit as the instanton-generated potential for the axion fields.  The coefficient $\gamma$ expresses the strength of a kinetic mixing~\cite{mp} between the axion field and the corresponding gravitational axion field, which, in four space-time dimensions, is dual to the aforementioned field strength of the antisymmetric tensor field  of the string gravitational multiplet.
For real $\gamma$,  
the approach is only valid for
\begin{align}\label{gammaregion}
\gamma < 1\ ,  \qquad \gamma \in {\tt R}, 
\end{align}
The mechanism for  the anomalous Majorana mass generation proposed in \cite{mp} leads to a two-loop  Majorana
right-handed neutrino $\psi_R$ mass: 
\begin{align}\label{massR}
M_R \sim \frac{\sqrt{3}\, \lambda\, \gamma\, c_1\, \kappa^8 \Lambda^9}{49152\sqrt{8}\,
\pi^4 (1 - \gamma^2 )}\; ,  
\end{align}
where $\Lambda$ is an Ultra-Violet (UV) momentum cutoff. In an UV 
complete theory,  such as  strings, $\Lambda$  and  $\kappa^{-1}$  are related~\cite{mp} via the string mass scale and compactification radii of the extra-dimensional spatial manifold. 
For a generic quantum gravity model, independent of string theory, one may use simply  $\Lambda \sim \kappa^{-1}$. As explained in \cite{mp}, the sterile fermion mass \eqref{massR} is {\it independent} of the axion-$a(x)$ potential, and thus its mass. 

In \cite{amsot} we have suggested that the constraint \eqref{gammaregion} can be evaded, 
upon the {\it simultaneous} complexification of the parameters $\gamma, \lambda$, which now become purely imaginary 
\begin{align}\label{complex}
\gamma \rightarrow i \tilde \gamma, \quad \lambda \rightarrow i \tilde \lambda, \qquad \tilde \gamma, \, \tilde \lambda \in {\tt R}.
\end{align}
In such a case one is effectively working with {\it non-Hermitean} Hamiltonians but connected to the so-called PT symmetric framework~\cite{bender}. We remark that in the case of axion (pseudoscalar) fields,  the Yukawa interaction term is PT-symmetry odd. On the other hand, 
if we had a scalar field in such interactions, one would obtain PT symmetric Hamiltonians~\cite{bender} of the type discussed in \cite{ptneutrino}, which have been argued to be consistent field theories describing phenomenologically relevant neutrino oscillations.  Nonetheless, as discussed in \cite{AM}, such interactions can lead to real energies in a certain regime of their parameters, thus making a connection with PT symmetric systems~\cite{BJR}.  In fact, according to the discussion in \cite{mann}, the existence of real energies is guaranteed if one has an underlying anti-linear symmetry, which could be more general than PT.  In our case such an antilinear symmetry is the CPT symmetry~\cite{AM}.  In \cite{AM} we also demonstrated the consistency of such non-Hermitian models with Lorentz invariance (including  also improper Lorentz transformations), as well as unitarity~\cite{AMS}.

It is important to note that, upon \eqref{complex}, the anomalous fermion mass \eqref{massR} remains  {\it real}:
\begin{align}\label{massR2}
M_R^{\rm anti-herm}\,  \sim \, - \frac{\sqrt{3}\, \tilde \lambda\, \tilde \gamma\, c_1\,  \kappa^8 \Lambda^9}{49152\sqrt{8}\,
\pi^4 (1 + \tilde \gamma^2 )}\; .
\end{align} 
Notice that the sign of the fermion mass depends on the sign  of the product $\tilde \lambda\, \tilde \gamma$, and can always be chosen to be positive, although for fermions, unlike bosons,  such a sign is not physically relevant. 

In \cite{amsot} we presented an alternative way to generate a sterile-neutrino mass,  dynamically induced by the Yukawa axion-sterile-fermion interaction via a study of the pertinent Schwinger-Dyson (SD) equations. We examined the generation 
of a dynamical mass for {\it both} fermion and (pseudo)scalar fields, in cases with hermitian and antihermitian Yukawa interactions. In \cite{amsot} we ignore anomalies, setting $\gamma=\tilde \gamma =0$. In the context of the action \eqref{bacoupl3}, then, we have shown that dynamical masses for the fermion and axion fields, $m$ and $M$ respectively, of approximately equal magnitude, is possible, provide there is a bare axion mass $M_0$:
\bea\label{SD1cb}
m^2 &\simeq&  \exp \Big(-\frac{16\, \pi^2}{\lambda^2}\Big) \, \Lambda^2, \quad |\lambda| \ll 1, \nonumber \\
M^2 &\simeq &  m^2 = M_0^2 - \frac{\lambda^2}{4\pi^2}\, \Lambda^2, \quad M_0^2 
= \frac{\lambda^2}{4\pi^2}\, \Lambda^2 + \exp \Big(-\frac{16\, \pi^2}{\lambda^2}\Big) \, \Lambda^2.
\eea
which indicates non-perturbative (in  the Yukawa coupling $\lambda$) small dynamical fermion and (pseudo)scalar masses. 

For the case of antihermitian Yukawa couplings \eqref{complex} (with $\tilde \gamma =0$), dynamical mass for the fermions is {\it not} possible, due to energetics~\cite{AM,VW}, but for (pseudo)scalar fields one can have a one-loop induced mass 
\begin{align}\label{axdynam}
(M^{(1)})^2 = \frac{\lambda^2}{4\pi^2}\Lambda^2 > 0, \, |\lambda | \ll 1, 
\end{align}
in the absence of a bare (pseudo)scalar mass $M_0=0$. 

In the presence of additional {\it attractive four-fermion} interactions in the Lagrangian of the model \eqref{bacoupl3}, say of the form~\footnote{Such four-fermion interactions may characterise the model of \cite{mp}, due to exchanges of massive mediator fields in the string-inspired effective  theory`\cite{amsot}.}
\be\label{fourf}
 - \frac{1}{2 f^2_4} (\overline{\psi} \, \gamma^5 \psi)^2,
 \ee
with $f_4$ a dimensionful coupling with mass dimension +1,  the situation changes drastically~\cite{amsot}. For
hermitian Yukawa couplings, dynamical masses for fermions and scalars  of order 
\begin{align}\label{4fmodm1bc}
m^2 & \simeq  M^2  \simeq  M_0^2  -  \frac{\lambda^2\, \Lambda^2}{4\pi^2} \,  
\simeq  \lambda^2 \, f_4^2 + \mathcal O(\lambda^4 \ln \lambda^2)
= \lambda^2 \, \frac{\Lambda^2}{16\pi^2} + \mathcal O(\lambda^4\,\ln \lambda^2) \ll \Lambda^2, \nonumber \\
M_0^2 &\simeq   \lambda^2 \, \frac{5\,\Lambda^2}{16\pi^2} + \mathcal O(\lambda^4 \,\ln \lambda^2)\,,
\end{align}
can be generated, which are {\it much larger} than the masses \eqref{SD1cb} in the pure Yukawa case where 
$f_4 \to \infty$. As follows from \eqref{4fmodm1bc}, we observe that, for consistency, the four-fermion coupling has to be proportional to the UV cutoff~\cite{amsot}, 
\be\label{4fcL}
f_4 \sim \frac{\Lambda}{4\pi}\,.
\ee
In the antihermitian Yukawa interaction case, in the presence of the interactions \eqref{fourf}, one can obtain dynamical fermion and (pseudo) scalar mass generation, in the absence of bare (pseudo)scalar mass $M_0=0$, which are of the same order as in the corresponding hermitian-Yukawa case \eqref{4fmodm1bc}:
\begin{align}
\label{nh4mu0ab}
 m^2 \simeq M^2 \simeq 4 \, \lambda^2 \, f_4^2  +  \Big | \mathcal O (\lambda^4 \, \ln \lambda^2) \Big | 
 \simeq   \frac{\lambda^2}{4 \pi^2} \Lambda^2  +  \Big | \mathcal O (\lambda^4 \, \ln \lambda^2) \Big |
\end{align}
with the four fermion coupling given by \eqref{4fcL}, as in the hermitian Yukawa interactions case.

The above results can be straightforwardly extended to the Majorana spinor case~\cite{amsot}, of direct relevance to the model of \cite{mp}, up to some numerical factors of 2. Specifically, for the corresponding case \eqref{SD1cb}, one has for the Majorana fermion dynamical mass:
\bea\label{SD1cbe}
m^2 &\simeq&  \exp \Big(-\frac{8\, \pi^2}{\lambda^2}\Big) \, \Lambda^2, \quad |\lambda| \ll 1, \nonumber \\
m^2 &\simeq& M^2 \simeq   M_0^2 - \frac{\lambda^2}{4\pi^2}\, \Lambda^2, \quad M_0^2 
= \frac{\lambda^2}{4\pi^2}\, \Lambda^2 + \exp \Big(-\frac{8\, \pi^2}{\lambda^2}\Big) \, \Lambda^2.
\eea
In the presence of four-fermion interactions \eqref{fourf}, 
the solution for dynamical fermion and scalar masses  $m \simeq M$ is qualitatively similar   as in 
\eqref{4fmodm1bc}, but with a different value of $f_4$:
\begin{align}\label{dfmmaj}
M^2 & \simeq  M_0^2  -  \frac{\lambda^2\, \Lambda^2}{4\pi^2} \,
\simeq  m^2 \simeq  \lambda^2 \, f_4^2 + \mathcal O(\lambda^4 \ln \lambda^2)~, \quad
f_4 \simeq \frac{\Lambda}{2\sqrt{2}\,\pi} + \Big| \mathcal O (\lambda^2\, \ln \lambda^2) \Big|, \quad \lambda^2 \ll 1~, \nonumber \\
M_0^2 &=\frac{3\, \lambda^2}{8\pi^2}\, \Lambda^2\,. 
\end{align}

In this article we shall examine the role of the anomaly 
coefficients $\gamma, \tilde \gamma$ in the appropriate hermitian or non-hermitian models. 
In the presence of such coefficients, radiative masses for the fermions are generated anomalously, 
as discussed in \cite{mp}. The point of the current work is to compare the dynamically generated mass with these radiative masses.
To ensure the reality of the radiative masses, and thus avoid quantum instabilities in the non-hermitian models, we impose the simultaneous complexification \eqref{complex}, which will determine the relevant cases.  

In addition to the anomaly terms, we also include a  ({\it constant}) axial background $\mathcal B_\mu$ 
term (where $\mathcal B_\mu$ has mass dimension +1). The inclusion of such a term  is motivated by CPT Violating models for leptogenesis involving heavy sterile neutrinos~\cite{decesare}. Indeed, the decay of the latter in the presence of  constant axial backgrounds of the form (violating spontaneously CPT and Lorentz symmetries)
\be\label{constant}
\mathcal B_{\mu} = \delta_{\mu\, 0}\, |\mathcal B|, \quad |\mathcal B| = {\rm constant}, \quad \mu=0, \dots 3,  
\ee
with $\delta_{\mu\,0}$ a Kronecker delta, can lead to lepton asymmetries at tree level. In the models of \cite{decesare}, the lagrangians are hermitian and bare masses for the sterile neutrinos have been assumed. It is the purpose of this section to examine whether the axial background itself can induce 
dynamical masses or the fermions and/or axions, which in turn could have implications for 
leptogenesis, as per the conclusions of \cite{decesare}. 

However, we shall also go beyond the hermitian cases. In particular we shall include antihermitian axial backgrounds $\mathcal B_\mu$. Such a case (but considering antihermitian axial backgrounds in a Nambu-Jona-Lasinio fermionic only model~\cite{NJL}, in the absence of Yukawa interactions, following a one-loop and not a complete SD treatment) has also been considered independently in \cite{klevansky} with the conclusion that the background enhances 
the dynamical mass generation induced by the four fermion interactions of the model. We shall confirm this result in our model as well, using a SD analysis~\cite{amsot}. 

We therefore consider the prototype Lagrangian (using \eqref{anomfermion} for the gravitational anomaly terms to express them in terms of axial fermion currents):
\begin{equation}\label{bacanomlag}
  \mathcal{L}= \frac{1}{2} \partial_{\mu} a \partial^{\mu} a + \bar{\psi} i \slashed{\partial} \psi - \frac{\gamma}{f_b}
  (\partial_{\mu} a) \bar{\psi} \gamma^{\mu} \gamma^5 \psi 
  +  \mathcal{B}_{\mu} \bar{\psi}
  \gamma^{\mu} \gamma^5 \psi 
  + i \lambda a \bar{\psi} \gamma^5 \psi - \frac{g^2}{2 f^2_b}
  (\bar{\psi} \gamma^5 \psi)^2\,,
\end{equation}
where, in the context of the model \cite{mp}, $f_b$ is given by \eqref{fb}, but of course one can treat it  more generally as an arbitrary mass scale to be determined self consistently in terms of the UV cutoff $\Lambda$ in the phase where dynamical mass generation occurs~\cite{amsot}. 
\nicktext{As we discussed in \cite{amsot}, in the context of the string-inspired models 
of \cite{mp}, one might assume a common origin of the coefficients $\gamma$ and $\lambda$, traced back to the same stringy non-perturbative effects deep in the stringy quantum gravity regime. This would imply $\gamma \sim \lambda$ and provide a natural explanation for the
appearance of a bare mass for the axions, necessary for dynamical mass generation of both fermions and axions (of the same order) in the hermitian Yukawa-interaction model. In the non hermitian-Yukawa-interaction case, the assumption on the equality of the magnitudes of $\gamma$ and $\lambda$ is not necessary, because  dynamical mass generation for axions occurs without the presence of a bare mass. Thus, in this work we consider $\gamma$ and $\lambda$ as independent parameters, which however could be linked for hermitian models.}
 
For concreteness we restrict our discussion to Dirac fermions, given that extension to the Majorana fermion case is straightforward and does not change qualitatively the conclusions. We shall return to the Majorana case when we compare the dynamical and radiatively anomalous  fermion masses. The reader should have noticed that in \eqref{bacanomlag}, we have introduced the 
generic notation $\gamma$ ($\lambda$) for the anomaly coefficients (Yukawa couplings) $\frac{\gamma}{\sqrt{1-\gamma^2}}$ ($\frac{\lambda}{\sqrt{1-\gamma^2}}$) for both hermitian and antihermitian cases, the latter obtained upon using \eqref{complex}. 
Moreover we used 
$g^2/f_b^2$ instead of $1/f_4^2$ ({\it cf.} \eqref{fourf})  for the coupling of the attractive four-fermion-interaction terms. This is for notational convenience, to make a more direct contact with the model \eqref{bacoupl3}~\cite{mp}, especially when we compare the contributions to  the dynamical fermion mass coming from the anomaly terms with those due to the four-fermion interactions, as well as when we make a comparison between the radiative ({\it cf.} \eqref{massR}, \eqref{massR2}) and dynamical fermion masses. 

At this point we stress that the axial current is classically conserved only in the case of massless chiral fermions, and in such a case its conservation might be spoiled by anomalies~\cite{anomalies}, {\it cf.} \eqref{anomfermion}. In the case of massive fermions, however, of mass m, one has classically that $\nabla_\mu J^{5\,\mu} = 2 \, i\, m \, \overline \psi \, \gamma^5 \, \psi$. Thus, when one considers the full quantum fermionic path integral, as necessary in SD treatments of dynamical fermion mass generation, the anomaly terms play in general a non-trivial role in mass generation, as we shall discuss in this work. 

Below we shall examine various cases involving both hermitian and antihermitian interactions, as per the initial motivations of this work, outlined in the introduction and in \cite{amsot}. We shall not deal with the most general solutions of the SD equations, but instead we shall  
extend appropriately our previous considerations to include axial background and anomaly terms. 
In particular, as far as the anomaly terms are concerned, we shall insist on maintaining the reality of the radiatively induced fermion masses \eqref{massR}, \eqref{massR2}, otherwise there would be instabilities in our system. This means that we shall consider cases in which 
the product
$\gamma \lambda $ always remains real 
\be\label{product}
\gamma \, \lambda \, \in \, \mathbb R \,,
\ee
that is, one should examine models with either $\gamma, \, \lambda  \in \mathbb R$ or 
$i\gamma, \, i\lambda  \in \mathbb R$. As already mentioned, for brevity,  we shall use the same symbols $\gamma$ and $\lambda$ in both hermitian and antihermitian cases, with the understanding though of the above restriction \eqref{product}. 

 Regarding the quantum SD treatment of antihermitian interactions, we should mention that in \cite{amsot} and here, we do not attempt a rigorous definition of the path integral measure of the various fields. Such an approach has been undertaken recently in \cite{bs}  in the context of a simplified  PT symmetric theory~\cite{bender} of a pseudoscalar field. In our case,  the situation is more complicated because of the Yukawa interactions of the pseudoscalar fields $a(x)$ with fermions, {\it cf. }\eqref{bacoupl3}. In our construction of the SD equations in \cite{amsot} and here, we use only formally the appropriate non-hermitian interactions, by working in a Euclidean path-integral,
and search for consistent solutions for mass generation for both (pseudo)scalar and fermion fields. 

In general, there are two types of masses for the fermions that can be generated dynamically, a Dirac mass, $m$, corresponding to a term in the effective action in the massive phase of the form 
$m \, \bar \psi \, \psi $, and a chiral mass $\mu $ corresponding to a term $\mu \, \bar \psi \, \gamma^5 \, \psi $. Despite the presence of antihermitian Yukawa interactions, such theories have real energies~\cite{BJR}, provided  $|m| > |\mu|$ (we stress again that, in the presence of antihermitian field-theoretic Yukawa interactions, the underlying anti-linear symmetry~\cite{mann} that guarantees the reality of the energies, under the above condition for the (dynamical) masses, is CPT symmetry~\cite{AM}). 
This motivated the inclusion of such non-hermitian models in our studies of dynamical mass generation in \cite{AM,amsot}, which was not considered in \cite{bender4f}. 
For our purposes here, we note that we seek solutions for dynamical mass generation for which the chiral mass, or equivalently a chiral condensate 
$<\bar \psi \, \gamma^5 \, \psi > $, vanishes $\mu=0$. Such solutions are consistent with considering 
real pseudoscalar fields in the path integral, as done in \cite{bs}, for which case  one would obtain from \eqref{bacoupl3} a classical equation of motion for the scalar fields of the form (including a real mass $M$ for the scalars, in general)
\be\label{eqscalar}
(\Box + M^2)\, \phi = i \lambda \, \bar \psi \, \gamma^5 \, \psi  +  \frac{\gamma}{f_b} \nabla_\mu (\bar \psi \, \gamma^\mu \, \gamma^5 \, \psi ) \,, 
\ee
corresponding to a saddle point in the path integral. We next renark that, in the absence of anomalies, the $\gamma$ term on the right-hand side of \eqref{eqscalar} vanishes for massless fermions. In the case of massive fermions, of mass $m$, this term reads simply $2i \, \frac{\gamma \, m}{f_b} \, \, \bar \psi \, \gamma^5 \, \psi $. 
For real (pseudo)scalar fields, in the antihermitian case \eqref{complex}, this equation would imply, for the general case $2\, \gamma \frac{m}{f_b} \ne -\lambda$ we consider here~\cite{amsot,mp}, 
the constraint~\cite{AM,amsot}: $\bar \psi \, \gamma^5 \, \psi = 0$.  This is associated with the fact that, for the saddle point, where quantum fluctuations of the fields are ignored, the left hand side of \eqref{eqscalar} is real, for real $\phi$ fields we consider here,  while the right-hand side is imaginary (it goes without saying, of course, that \eqref{eqscalar} is not satisfied by the quantum fields away from the saddle point, for which there are no restrictions). 
For our purposes, it is sufficient that we find consistent solutions with the above constraint implemented dynamically, as a vanishing condensate for the fermion fields, $\langle \bar \psi \, \gamma^5 \, \psi  \rangle = 0$, or equivalently $\mu=0$. Energetics arguments~\cite{AM,VW} support this point of view in the case of antihermitian Yukawa interaction models. In \cite{amsot} we extended this constraint also to the models with hermitian Yukawa interactions, and this will also characterise the SD analysis in the present article, in the sense of setting the chiral mass $\mu=0$ in all our SD equations, looking for non-trivial solutions for dynamically generated Dirac mass $m \ne 0$.

The structure of the article is as follows: in the next section \ref{sec:habhy}, we discuss SD dynamical mass generation in the presence of hermitian axial background and hermitian Yukawa interactions, in the absence of anomaly terms, both in the presence and absence of four fermion attractive interactions. 
This is because we want to study the effects of the axial background prior to inclusion of non-trivkal anomaly coefficients. In section \ref{sec:habnhy}, we repeat the analysis for hermitian axial backgrounds but antihermitian Yukawa interactions, again in the absence of anomalies. In section \ref{sec:nhabhy} we study non hermitian axial backgrounds and hermitian Yukawa interactions, while in 
section \ref{sec:nhabnhy} we complete our study on the role of axial backgrounds in mass generation for non-anomalous models by examining  the case of antihermitian axial backgrounds  in the presence of antihermitian Yukawa interactions. In section \ref{sec:hahy}, we discuss the role of hermitian anomaly terms in dynamical mass generation induced by hermitian Yukawa interactions, in the absence of axial backgrounds, while in section \ref{sec:nhanhy} we repeat the study for the case where the anomaly and the Yukawa interaction terms are both antihermitian. Finally, section \ref{sec:concl} contains our conclusions and outlook, as well as a comparison between the dynamically generated fermion masses and the radiative (anomalously-generated) ones \eqref{massR}, \eqref{massR2}, in the context of the sterile neutrino model of \cite{mp}.

\section{Hermitian Axial Backgrounds and Hermitian Yukawa Interactions - no anomalies\label{sec:habhy}} 

In this and the subsequent sections we shall be brief with technical details on the SD formalism, given that those have been provided in \cite{amsot}, where we refer the interested reader for details. We shall concentrate instead in merely giving the final SD equations for mass generation and describing the pertinent solutions. 

Let us first ignore the anomaly coefficient, by setting $\gamma=0$. 
In this case, the Lagrangian \eqref{bacanomlag} becomes:
\begin{equation}
\label{backlaginit}
  \mathcal{L}= \frac{1}{2} \partial_{\mu} a \partial^{\mu} a + \bar{\psi} i
  \slashed{\partial} \psi + \mathcal{B}_{\mu} \bar{\psi}
  \gamma^{\mu} \gamma^5 \psi + i \lambda a \bar{\psi} \gamma^5 \psi -
  \frac{g^2}{2 f^2_b} (\bar{\psi} \gamma^5 \psi)^2, \quad \gamma, \lambda \, \in \, \mathbb R.
\end{equation}

When the background term $\mathcal{B}_{\mu}$ is constant, one can incorporate it in the fermionic propagator. Upon linearising the four fermion interaction using the auxiliary field $\sigma$, 
we write for the generating functional
\begin{eqnarray}
  Z [K, J, \eta, \bar{\eta}] &=& \int \mathcal{D} [\sigma a \psi \bar{\psi}]
  \exp \left\{ i \int d^4 x \left[ \frac{1}{2} \partial_{\mu} a \partial^{\mu}
  a + \bar{\psi} \left( i \slashed{\partial} + \mathcal{B}_{\mu}
  \gamma^{\mu} \gamma^5 \right) \psi + i \lambda a \bar{\psi}
  \gamma^5 \psi - \frac{f^2_b}{2 g^2} \sigma^2 - i \sigma \bar{\psi} \gamma^5
  \psi \right] \right.\nonumber\\
  & & \left. + i \int d^4 x [J a + \bar{\psi} \eta + \bar{\eta} \psi + K
  \sigma] \right\}
\end{eqnarray}
We note, for the benefit of the reader, that the source terms,  including that ($K$) for the auxiliary field $\sigma$, are introduced for the consistent derivation of the SD equations~\cite{amsot}, which is not given here, for brevity. 

In this way, the interaction part  is the same as in the hermitian Yukawa interaction model, 
in the presence of attractive four-fermion interactions, 
discussed in \cite{amsot}. Therefore, the corresponding SD equations are formally the same as in that case, with the difference that now the fermionic propagator contains the additional background  $\mathcal B_\mu$ terms. Following the argumentation in \cite{amsot}, which was also reviewed briefly at the end of the introductory section \ref{sec:intro}, we look for dynamical mass scenarios in which the chiral fermion mass vanishes, $\mu =0$ and only a Dirac mass $m$ is generated. This will characterise all the cases considered in the current work. 

In the rainbow approximation sufficient for weak Yukawa interactions (with $|\lambda | \ll 1$) we are restricting ourselves here and in \cite{amsot},\footnote{In this work, as in \cite{amsot}, we ignore wave function renormalization and vertex corrections, which are sufficient and self consistent approximations for our case of small Yukawa 
couplings. Such corrections will only lead to improvements which however will not change qualitatively our results. In this respect, we mention that our models, based on \eqref{bacanomlag}, are different from the model of \cite{bashir} with scalar self interactions, for which wave function renormalization, and the associated Ward-Takahashi identities, are important in yielding dynamical fermion mass above a 
critical value of the (common) self- and Yukawa-  interactions coupling in that model. In our case~\cite{amsot}, in the scalar sector we allow at most a bare scalar mass, and we look for non-trivial solutions 
for both (peudo)scalar and fermion mass generation,  in a self-consistent way,  
for {\it small} Yukawa couplings.} the pertinent SD equations, in the presence of a bare scalar mass $M_0$, whose introduction is necessary in this case~\cite{amsot} in order to have dynamical mass for the scalar field, read:

\begin{equation}
\label{backets1}
  4 i m = - \lambda^2  \tmop{tr} \left( \int_p \gamma^5
  G_f (p) \gamma^5 G_s (p) \right) - \tmop{tr} \left(\gamma^5 \int_p G_f (p) \gamma^5
  G_{\sigma} (p) \right)
\end{equation}
\begin{equation}
\label{backets2}
  i M^2 = i M_0^2 + \lambda^2 \tmop{tr} \left[ \int_p \gamma^5 G_f (p)
  \gamma^5 G_f (p) \right]
\end{equation}
where the SD dressed propagators for the fermion field in the (constant) axial background ($G_f$), the (pseudo)scalar field  ($G_s$) 
and the auxiliary scalar $\sigma$ field that linearises the four-fermion interactions ($G_\sigma$), 
are given by~\cite{amsot}:\\
$G_f (p) = \frac{i}{\slashed{p} + \mathcal{B}_{\mu} \gamma^{\mu} \gamma^5 - m}$, $G_s=\frac{i}{p^2-M^2}$, and $G_\sigma (p) = -\frac{i\, g^2}{f_b^2}$, where $M$ and $m$ denote the dynamical scalar and Dirac-fermion masses,  respectively, and we use 
the notation $\slashed{p} = \gamma^\mu p_\mu$, with $\gamma^\mu$ the Dirac $\gamma$-matrices,  and $p^2 = p_\mu p^\mu$.

To solve the SD equations we follow a similar treatment as in \cite{klevansky}, by writing the fermion propagator as

\begin{equation}
\label{ferpropag}
\frac{1}{\slashed{p} + \mathcal{B}_{\mu} \gamma^{\mu} \gamma^5 - m}
= \frac{\left( \slashed{p} -  \mathcal{B}_{\mu} \gamma^{\mu} \gamma^5
+ m \right) \left( \left( p^2 + \mathcal{B}^2 - m^2 \right) +
2  \mathcal{B} \cdot p \gamma^5 - 2 m 
\mathcal{B}_{\mu} \gamma^{\mu} \gamma^5 \right)}{p^4 + 2 \left(
 \mathcal{B}^2 - m^2 \right) p^2 - 4 
(\mathcal{B} \cdot p)^2 + \left( \mathcal{B}^2 + m^2
\right)^2}.
\end{equation} 
After Wick rotation in (\ref{backets1}) and (\ref{backets2}), use of the spherical coordinates $(p,\theta,\phi_1,\phi_2)$, so as to write $\mathcal{B} \cdot p=|\mathcal{B}| |p| \cos{\theta}$, and using an UV  cutoff $\Lambda$ in the ``radial'' integral, we obtain the following system of algebraic equations:
\begin{eqnarray}
\label{sdsolbackscal}
  M^2 &=& M_0^2 - \frac{\lambda^2}{16 \pi^2 \mathcal{B}^2}
  \left( \Lambda^2 \left( \Lambda^2 + 2 \alpha \beta - \sqrt{(\alpha^2 + \Lambda^2) (\beta^2 +
  \Lambda^2)} \right) + \frac{1}{2} \left( \sqrt{(\alpha^2 + \Lambda^2) (\beta^2 +
  \Lambda^2)} - | \alpha | \beta \right) (3 (\alpha^2 + \beta^2) - 8 \alpha \beta) \right.\nonumber\\
  & & \left. - \frac{1}{2} (\alpha -
  \beta)^2 (3 (\alpha^2 + \beta^2) - 2 \alpha \beta) \ln \left[ \frac{\sqrt{(\alpha^2 + \Lambda^2)} +
  \sqrt{(\beta^2 + \Lambda^2)}}{| \alpha | + \beta} \right] \right)
\end{eqnarray}
and

\begin{eqnarray}
\label{sdsolbackferm}
  1 &=& \frac{\lambda^2}{32 \pi^2 \mathcal{B}^2} \left(
  \Lambda^2 - (M^2 - \alpha \beta) \left( 1 + \frac{M^2 - \alpha \beta}{\sqrt{(\alpha^2 - M^2) (\beta^2 -
  M^2)}} \right) \ln \left( 1 + \frac{\Lambda^2}{M^2} \right) - \left(
  \sqrt{(\alpha^2 + \Lambda^2) (\beta^2 + \Lambda^2)} - | \alpha | \beta \right) \right.\nonumber\\
  & & \left. + ((\alpha - \beta)^2 +
  2 (M^2 - \alpha \beta)) \ln \left[ \frac{\sqrt{(\alpha^2 + \Lambda^2)} + \sqrt{(\beta^2 +
  \Lambda^2)}}{| \alpha | + \beta} \right] \right.\nonumber\\
  & & \left. + \frac{(M^2 - \alpha \beta)^2}{\sqrt{(\alpha^2 - M^2)
  (\beta^2 - M^2)}} \ln \left[ \frac{\left( \sqrt{(\alpha^2 - M^2) (\beta^2 - M^2)} +
  \sqrt{(\alpha^2 + \Lambda^2) (\beta^2 + \Lambda^2)} \right)^2 - (M^2 +
  \Lambda^2)^2}{\left( | \alpha | \beta + \sqrt{(\alpha^2 - M^2) (\beta^2 - M^2)} \right)^2 -
  M^4} \right] \right) \nonumber\\
  & & + \frac{g^2}{64 \pi^2 f_b^2 \mathcal{B}^2} \left( \Lambda^2
  \left( \Lambda^2 + 2 \alpha \beta - \sqrt{(\alpha^2 + \Lambda^2) (\beta^2 + \Lambda^2)}
  \right) + \frac{1}{2} \left( \sqrt{(\alpha^2 + \Lambda^2) (\beta^2 + \Lambda^2)} - |
  \alpha | \beta \right) (3 (\alpha^2 + \beta^2) - 8 \alpha \beta) \right.\nonumber\\
  & & \left. - \frac{1}{2} (\alpha - \beta)^2 (3 (\alpha^2 + \beta^2)
  - 2 \alpha \beta) \ln \left[ \frac{\sqrt{(\alpha^2 + \Lambda^2)} + \sqrt{(\beta^2 +
  \Lambda^2)}}{| \alpha | + \beta} \right] \right)
\end{eqnarray}

Where we have defined $\alpha = m - | \mathcal{B} |$ and $\beta = m + |\mathcal{B} |$.

\subsection{Dynamical (pseudo)scalar mass}

To see whether the axial background $\mathcal B_\mu$ is capable of inducing a dynamical mass for the (pseudo)scalar field, we 
set the bare mass to zero $M_0=0$. We shall consider the case $m \simeq M$.
On substituting in \eqref{sdsolbackscal}, then, and defining $\bar{M}=M/\Lambda$ and $\bar{\mathcal{B}}=\mathcal{B}/\Lambda$ we obtain:

\begin{eqnarray}
\label{nomass}
  1 &=& - \frac{\lambda^2}{16 \pi^2 \bar{M}^2
  \bar{\mathcal{B}}^2} \left( 1 + 2 (\bar{M}^2 - \bar{\mathcal{B}}^2) -
  \sqrt{(1 + \bar{M}^2 + \bar{\mathcal{B}}^2)^2 - 4 \bar{M}^2
  \bar{\mathcal{B}}^2} \right.\nonumber\\
  & & \left. - \left( \sqrt{(1 + \bar{M}^2 + \bar{\mathcal{B}}^2)^2
  - 4 \bar{M}^2 \bar{\mathcal{B}}^2} - \sqrt{(\bar{M}^2 -
  \bar{\mathcal{B}}^2)^2} \right) (\bar{M}^2 - 7 \bar{\mathcal{B}}^2) \right.\nonumber\\
  & & \left. - 4 \bar{\mathcal{B}}^2 (\bar{M}^2 + 2 \bar{\mathcal{B}}^2) \log \left[ \frac{1
  + \bar{M}^2 + \bar{\mathcal{B}}^2 + \sqrt{(1 + \bar{M}^2 +
  \bar{\mathcal{B}}^2)^2 - 4 \bar{M}^2 \bar{\mathcal{B}}^2}}{(\bar{M}^2 +
  \bar{\mathcal{B}}^2) + \sqrt{(\bar{M}^2 - \bar{\mathcal{B}}^2)^2}} \right]
  \right)
\end{eqnarray}

\begin{figure}[ht]
 \centering
  \includegraphics[clip,width=0.65\textwidth,height=0.25\textheight]{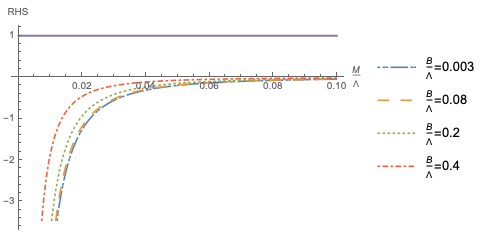} 
\caption{\it The Plot shows different values of the right hand side (r.h.s.) of \eqref{nomass} for the fixed value $\lambda^2=0.02$.  
There is no intersection with unity, which means that, for Hermitian Yukawa coupling and hermitian constant axial background,  there is no dynamical mass generation for the scalar, in the case where the bare scalar mass vanishes, $M_0=0$. When $M_0 \ne 0$, the r.h.s. of \eqref{nomass} 
becomes $1 - \frac{M_0^2}{M^2}$, with $M_0$ to be determined, after the dynamical 
fermion mass is determined through Eq.~\eqref{sdsolbackferm}.}\label{fig:plotback1}
\end{figure}

 We plot the right hand side of \eqref{nomass} for different values of $\mathcal{B}$ in fig.~\ref{fig:plotback1}. As we can readily see from the figure, this function never intersects 1, 
which means that there is no solution, that is the background cannot induce dynamically a scalar mass
in the absence of a bare mass $M_0$.  

Once a non zero bare mass $M_0 \ne 0$ is introduced, the latter can be determined by the fermion-mass SD equation \eqref{sdsolbackferm}, which depends only on the renormalised masses $m, M$. The bare mass $M_0$ is easily found from the small $|\mathcal{B}|$ limit of \eqref{sdsolbackscal}, after setting $m \simeq M$, under the condition $\Lambda^2 \gg M^2$: 

\begin{eqnarray}
\label{serieback2}
M_0^2 = M^2 + \frac{\lambda^2}{4 \pi^2} \Lambda^2 +
\frac{\lambda^2 \mathcal{B}^2}{8 \pi^2} \left[
7- 4 \ln \left(\frac{\Lambda^2}{M^2} \right) \right],
\end{eqnarray}
where $M$ can be determined from the solution to the fermionic SD equation, which we now proceed to discuss for the case $m \simeq M $. 

\subsection{Dynamical fermion mass}

On setting $m=M$ and defining $\bar{M}=M/\Lambda$, $\bar{\mathcal{B}}=\mathcal{B}/\Lambda$ and $\bar{g}=(g\Lambda)/f_b$, Eq.~\eqref{sdsolbackferm} becomes:
\begin{eqnarray}
\label{fermarr}
  1 &=& \frac{\lambda^2}{32 \pi^2 \bar{\mathcal{B}}^2} \left( 1
  - \bar{\mathcal{B}}^2 \left( 1 +
  \frac{\bar{\mathcal{B}}^2}{\sqrt{\bar{\mathcal{B}}^4 - 4 \bar{M}^2
  \bar{\mathcal{B}}^2}} \right) \log \left( 1 + \frac{1}{\bar{M}^2} \right) -
  \left( \sqrt{(1 + \bar{M}^2 + \bar{\mathcal{B}}^2)^2 - 4 \bar{M}^2
  \bar{\mathcal{B}}^2} - \sqrt{(\bar{M}^2 - \bar{\mathcal{B}}^2)^2} \right) \right.\nonumber\\
  & & \left. + 3 \bar{\mathcal{B}}^2 \log \left[ \frac{1 + (\bar{M}^2 +
  \bar{\mathcal{B}}^2) + \sqrt{(1 + \bar{M}^2 + \bar{\mathcal{B}}^2)^2 - 4
  \bar{M}^2 \bar{\mathcal{B}}^2}}{(\bar{M}^2 + \bar{\mathcal{B}}^2) +
  \sqrt{(\bar{M}^2 - \bar{\mathcal{B}}^2)^2}} \right] \right.\nonumber\\ 
  & & \left. + \frac{\bar{\mathcal{B}}^4}{\sqrt{\bar{\mathcal{B}}^4 - 4 \bar{M}^2
  \bar{\mathcal{B}}^2}} \log \left[ \frac{\left( \sqrt{\bar{\mathcal{B}}^4 - 4
  \bar{M}^2 \bar{\mathcal{B}}^2} + \sqrt{(1 + \bar{M}^2 +
  \bar{\mathcal{B}}^2)^2 - 4 \bar{M}^2 \bar{\mathcal{B}}^2} \right)^2 - (1 +
  \bar{M}^2)^2}{\left( \sqrt{(\bar{M}^2 - \bar{\mathcal{B}}^2)^2} +
  \sqrt{\bar{\mathcal{B}}^4 - 4 \bar{M}^2 \bar{\mathcal{B}}^2} \right)^2 -
  \bar{M}^4} \right] \right) \nonumber\\
  & & + \frac{\bar{g}^2}{64 \pi^2
  \bar{\mathcal{B}}^2} \left( 1 + 2 (\bar{M}^2 - \bar{\mathcal{B}}^2) -
  \sqrt{(1 + \bar{M}^2 + \bar{\mathcal{B}}^2)^2 - 4 \bar{M}^2
  \bar{\mathcal{B}}^2} \right.\nonumber\\
  & & \left. - \left( \sqrt{(1 + \bar{M}^2 + \bar{\mathcal{B}}^2)^2
  - 4 \bar{M}^2 \bar{\mathcal{B}}^2} - \sqrt{(\bar{M}^2 -
  \bar{\mathcal{B}}^2)^2} \right) (\bar{M}^2 - 7 \bar{\mathcal{B}}^2) \right.\nonumber\\
  & & \left. - 4 \bar{\mathcal{B}}^2 (\bar{M}^2 + 2 \bar{\mathcal{B}}^2) \log \left[ \frac{1
  + (\bar{M}^2 + \bar{\mathcal{B}}^2) + \sqrt{(1 + \bar{M}^2 +
  \bar{\mathcal{B}}^2)^2 - 4 \bar{M}^2 \bar{\mathcal{B}}^2}}{(\bar{M}^2 +
  \bar{\mathcal{B}}^2) + \sqrt{(\bar{M}^2 - \bar{\mathcal{B}}^2)^2}} \right]
  \right)
\end{eqnarray}

\begin{figure}[ht]
 \centering
  \includegraphics[clip,width=0.65\textwidth,height=0.25\textheight]{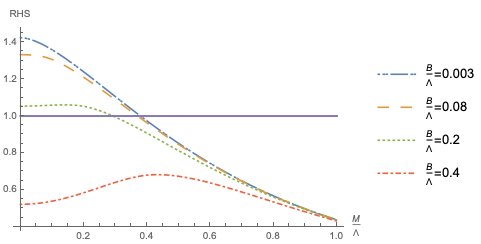} 
\caption{\it The Plot shows different values for the right hand side of \eqref{fermarr} for the fixed values $\lambda^2=0.02$ and $\bar{g}=15$. The continuous constant line (at the fixed value 1) represents the left hand side of Eq.~\eqref{fermarr}. The points at which the curves in the figure intersect this continuous line  correspond to solutions of \eqref{fermarr} yielding dynamical mass for the fermion.}\label{fig:plotbackfer}
\end{figure}

In fig.~\ref{fig:plotbackfer} we plot the curves corresponding to the right hand side of \eqref{fermarr} for different values of $\bar{\mathcal{B}}$. We notice that there are curves intersecting the constant solid line at 1, which corresponds to the left hand side of \eqref{fermarr},  for specific choices of $\lambda$ and $g$. For these cases we will have solutions for the equation \eqref{fermarr} with $m\simeq M$. We observe that there are solutions when the background field $\mathcal{B}$ becomes sufficiently small, $|\mathcal B| < \Lambda$,  corresponding to $m \simeq M < \Lambda$. 

At this stage we remark, that in the absence of the attractive four-fermion interactions, i.e. when $g=\bar g=0$,  there is dynamical fermion mass, for sufficiently small $|\mathcal B| < m$, 
which matches smoothly the non-perturbative solution found in \cite{amsot}, for $|\mathcal B| \to 0$.
This can be seen analytically by considering the small-$|\mathcal B|$ limit of the SD equation for the fermion, \eqref{fermarr}, which, up to and including order $|\mathcal B|^2$ terms, reads for $g=0$:
\begin{eqnarray}
\label{msimback}
1 = \frac{\lambda^2}{16 \pi^2}\, \ln \left[ 1 +
\frac{\Lambda^2}{M^2} \right] + \frac{\lambda^2 \mathcal{B}^2
\Lambda^4 (M^2 - \Lambda^2)}{32 \pi^2 M^2 (M^2 + \Lambda^2)^3} 
\simeq  \frac{\lambda^2}{16 \pi^2} \, \ln\left[\frac{\Lambda^2}{M^2} \right]  -  \frac{\lambda^2 \mathcal{B}^2}{32 \pi^2 M^2}, \quad |\lambda| \ll 1,
\end{eqnarray}
where the last equality is valid upon making the approximation $\Lambda \gg M \gg |\mathcal B|$. 
Equation \eqref{msimback} matches smoothly the situation for the hermitian Yukawa interaction described in \cite{amsot} in the limit $|\mathcal B| \to 0$. For small but finite values of $|\mathcal B| \ll M$
one obtains slightly modified non perturbative solutions, which are obtained from
\be\label{SDB0}
\frac{16\pi^2}{\lambda^2} = \ln\left[\frac{\Lambda^2}{M^2} \right]  - \frac{|\mathcal B|^2}{2\, M^2} , \quad 
\frac{|\mathcal B|^2}{2\, M^2} \ll 1.
\ee
The latter equation cannot be solved analytically but it is easy to see that there are solutions, sufficiently 
close to those found in the corresponding case studied in \cite{amsot}. Indeed, if we represent $\Lambda^2/M^2 = e^\xi, \, \xi \gg 1 > 0$, then we obtain from  \eqref{SDB0} the consistency condition $\xi - \frac{16\pi^2}{\lambda^2} = \frac{|\mathcal B|^2}{2\, M^2} \ll 1$, for $|\lambda| \ll 1$;  this has solutions for $\xi$, which turns out to take on a slightly higher value than the corresponding one 
in the absence of the axial background, leading to {\it a further suppression} of the dynamical fermion mass. The larger the magnitude of the background the bigger the suppression of the dynamical mass.

\section{Hermitian Axial Backgrounds and Non-Hermitian Yukawa Interactions - no anomalies \label{sec:habnhy}}

Next we study the case in which the Yukawa coupling is non-hermitian. For brevity, 
we shall omit the details of the derivation of the SD equations, and  give, instead, directly the solutions, which are 

\begin{eqnarray}
\label{sdsolbackynhscal}
  M^2 &=& \frac{\lambda^2}{16 \pi^2 \mathcal{B}^2}
  \left( \Lambda^2 \left( \Lambda^2 + 2 \alpha \beta - \sqrt{(\alpha^2 + \Lambda^2) (\beta^2 +
  \Lambda^2)} \right) + \frac{1}{2} \left( \sqrt{(\alpha^2 + \Lambda^2) (\beta^2 +
  \Lambda^2)} - | \alpha | \beta \right) (3 (\alpha^2 + \beta^2) - 8 \alpha \beta) \right.\nonumber\\
  & & \left. - \frac{1}{2} (\alpha -
  \beta)^2 (3 (\alpha^2 + \beta^2) - 2 \alpha \beta) \ln \left[ \frac{\sqrt{(\alpha^2 + \Lambda^2)} +
  \sqrt{(\beta^2 + \Lambda^2)}}{| \alpha | + \beta} \right] \right)
\end{eqnarray}
and

\begin{eqnarray}
\label{sdsolbackynhferm}
  1 &=& -\frac{\lambda^2}{32 \pi^2 \mathcal{B}^2} \left(
  \Lambda^2 - (M^2 - \alpha \beta) \left( 1 + \frac{M^2 - \alpha \beta}{\sqrt{(\alpha^2 - M^2) (\beta^2 -
  M^2)}} \right) \ln \left( 1 + \frac{\Lambda^2}{M^2} \right) - \left(
  \sqrt{(\alpha^2 + \Lambda^2) (\beta^2 + \Lambda^2)} - | \alpha | \beta \right) \right.\nonumber\\
  & & \left. + ((\alpha - \beta)^2 +
  2 (M^2 - \alpha \beta)) \ln \left[ \frac{\sqrt{(\alpha^2 + \Lambda^2)} + \sqrt{(\beta^2 +
  \Lambda^2)}}{| \alpha | + \beta} \right] \right.\nonumber\\
  & & \left. + \frac{(M^2 - \alpha \beta)^2}{\sqrt{(\alpha^2 - M^2)
  (\beta^2 - M^2)}} \ln \left[ \frac{\left( \sqrt{(\alpha^2 - M^2) (\beta^2 - M^2)} +
  \sqrt{(\alpha^2 + \Lambda^2) (\beta^2 + \Lambda^2)} \right)^2 - (M^2 +
  \Lambda^2)^2}{\left( | \alpha | \beta + \sqrt{(\alpha^2 - M^2) (\beta^2 - M^2)} \right)^2 -
  M^4} \right] \right) \nonumber\\
  & & + \frac{g^2}{64 \pi^2 f_b^2 \mathcal{B}^2} \left( \Lambda^2
  \left( \Lambda^2 + 2 \alpha \beta - \sqrt{(\alpha^2 + \Lambda^2) (\beta^2 + \Lambda^2)}
  \right) + \frac{1}{2} \left( \sqrt{(\alpha^2 + \Lambda^2) (\beta^2 + \Lambda^2)} - |
  \alpha | \beta \right) (3 (\alpha^2 + \beta^2) - 8 \alpha \beta) \right.\nonumber\\
  & & \left. - \frac{1}{2} (\alpha - \beta)^2 (3 (\alpha^2 + \beta^2)
  - 2 \alpha \beta) \ln \left[ \frac{\sqrt{(\alpha^2 + \Lambda^2)} + \sqrt{(\beta^2 +
  \Lambda^2)}}{| \alpha | + \beta} \right] \right)\,,
\end{eqnarray}
where again $\alpha = m - | \mathcal{B} |$ and $\beta = m + |\mathcal{B} |$. We have not incorporated 
a bare mass here, because, as we shall see below, a scalar mass can be dynamically generated
by the Yukawa coupling in this case. This is the main difference from the hermitian Yukawa coupling case. As we shall see below, the fermion mass exhibits a similar behavior with the background 
$|\mathcal B|$ as in the hermitian case. 

\subsection{Dynamical (pseudo)scalar mass \label{sec:hbsm}}

We are interested in solutions for $m \simeq M$. 
On setting $m\simeq M$, and defining $\bar{M}=M/\Lambda$ and $\bar{\mathcal{B}}=\mathcal{B}/\Lambda$ we obtain from \eqref{sdsolbackynhscal}

\begin{eqnarray}
\label{scalynh}
  1 &=& \frac{\lambda^2}{16 \pi^2 \bar{M}^2
  \bar{\mathcal{B}}^2} \left( 1 + 2 (\bar{M}^2 - \bar{\mathcal{B}}^2) -
  \sqrt{(1 + \bar{M}^2 + \bar{\mathcal{B}}^2)^2 - 4 \bar{M}^2
  \bar{\mathcal{B}}^2} \right.\nonumber\\
  & & \left. - \left( \sqrt{(1 + \bar{M}^2 + \bar{\mathcal{B}}^2)^2
  - 4 \bar{M}^2 \bar{\mathcal{B}}^2} - \sqrt{(\bar{M}^2 -
  \bar{\mathcal{B}}^2)^2} \right) (\bar{M}^2 - 7 \bar{\mathcal{B}}^2) \right.\nonumber\\
  & & \left. - 4 \bar{\mathcal{B}}^2 (\bar{M}^2 + 2 \bar{\mathcal{B}}^2) \log \left[ \frac{1
  + \bar{M}^2 + \bar{\mathcal{B}}^2 + \sqrt{(1 + \bar{M}^2 +
  \bar{\mathcal{B}}^2)^2 - 4 \bar{M}^2 \bar{\mathcal{B}}^2}}{(\bar{M}^2 +
  \bar{\mathcal{B}}^2) + \sqrt{(\bar{M}^2 - \bar{\mathcal{B}}^2)^2}} \right]
  \right)
\end{eqnarray}

\begin{figure}[ht]
 \centering
  \includegraphics[clip,width=0.65\textwidth,height=0.25\textheight]{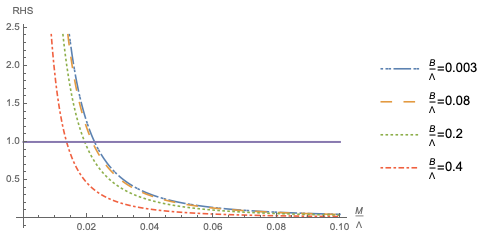} 
\caption{\it The Plot shows different values of the right hand side (r.h.s.) of \eqref{scalynh} for the fixed value $\lambda^2=0.02$.  
All the curves intersect with the constant curve at 1,  representing the left-hand side of \eqref{scalynh}, which means that, for non-Hermitian Yukawa coupling and hermitian constant axial background,  there is dynamical mass generation for the scalar, without the need to introduce a bare mass.}\label{fig:plotbacksynh}
\end{figure}

We plot in fig.~\ref{fig:plotbacksynh} the right hand side of the above equation. The reader can readily observe that the curves always intersect the solid line corresponding to the value $1$, representing the left hand side of Eq.~\eqref{scalynh}, 
which implies the existence of non-trivial  solutions to this equation. Thus, both fermion and scalar masses, of equal magnitude,  are dynamically generated in this case, in a similar spirit to the situation when the  axial background is absent. The presence of the background affects the magnitude of the mass. The larger the background, the smaller the dynamical mass.

\subsection{Dynamical fermion mass}

As before, using  $m\simeq M$ and defining $\bar{M}=M/\Lambda$ and $\bar{\mathcal{B}}=\mathcal{B}/\Lambda$, one obtains from  \eqref{sdsolbackynhferm}: 

\begin{eqnarray}
\label{fermarrynh}
  1 &=& -\frac{\lambda^2}{32 \pi^2 \bar{\mathcal{B}}^2} \left( 1
  - \bar{\mathcal{B}}^2 \left( 1 +
  \frac{\bar{\mathcal{B}}^2}{\sqrt{\bar{\mathcal{B}}^4 - 4 \bar{M}^2
  \bar{\mathcal{B}}^2}} \right) \log \left( 1 + \frac{1}{\bar{M}^2} \right) -
  \left( \sqrt{(1 + \bar{M}^2 + \bar{\mathcal{B}}^2)^2 - 4 \bar{M}^2
  \bar{\mathcal{B}}^2} - \sqrt{(\bar{M}^2 - \bar{\mathcal{B}}^2)^2} \right) \right.\nonumber\\
  & & \left. + 3 \bar{\mathcal{B}}^2 \log \left[ \frac{1 + (\bar{M}^2 +
  \bar{\mathcal{B}}^2) + \sqrt{(1 + \bar{M}^2 + \bar{\mathcal{B}}^2)^2 - 4
  \bar{M}^2 \bar{\mathcal{B}}^2}}{(\bar{M}^2 + \bar{\mathcal{B}}^2) +
  \sqrt{(\bar{M}^2 - \bar{\mathcal{B}}^2)^2}} \right] \right.\nonumber\\ 
  & & \left. + \frac{\bar{\mathcal{B}}^4}{\sqrt{\bar{\mathcal{B}}^4 - 4 \bar{M}^2
  \bar{\mathcal{B}}^2}} \log \left[ \frac{\left( \sqrt{\bar{\mathcal{B}}^4 - 4
  \bar{M}^2 \bar{\mathcal{B}}^2} + \sqrt{(1 + \bar{M}^2 +
  \bar{\mathcal{B}}^2)^2 - 4 \bar{M}^2 \bar{\mathcal{B}}^2} \right)^2 - (1 +
  \bar{M}^2)^2}{\left( \sqrt{(\bar{M}^2 - \bar{\mathcal{B}}^2)^2} +
  \sqrt{\bar{\mathcal{B}}^4 - 4 \bar{M}^2 \bar{\mathcal{B}}^2} \right)^2 -
  \bar{M}^4} \right] \right) \nonumber\\
  & & + \frac{\bar{g}^2}{64 \pi^2
  \bar{\mathcal{B}}^2} \left( 1 + 2 (\bar{M}^2 - \bar{\mathcal{B}}^2) -
  \sqrt{(1 + \bar{M}^2 + \bar{\mathcal{B}}^2)^2 - 4 \bar{M}^2
  \bar{\mathcal{B}}^2} \right.\nonumber\\
  & & \left. - \left( \sqrt{(1 + \bar{M}^2 + \bar{\mathcal{B}}^2)^2
  - 4 \bar{M}^2 \bar{\mathcal{B}}^2} - \sqrt{(\bar{M}^2 -
  \bar{\mathcal{B}}^2)^2} \right) (\bar{M}^2 - 7 \bar{\mathcal{B}}^2) \right.\nonumber\\
  & & \left. - 4 \bar{\mathcal{B}}^2 (\bar{M}^2 + 2 \bar{\mathcal{B}}^2) \log \left[ \frac{1
  + (\bar{M}^2 + \bar{\mathcal{B}}^2) + \sqrt{(1 + \bar{M}^2 +
  \bar{\mathcal{B}}^2)^2 - 4 \bar{M}^2 \bar{\mathcal{B}}^2}}{(\bar{M}^2 +
  \bar{\mathcal{B}}^2) + \sqrt{(\bar{M}^2 - \bar{\mathcal{B}}^2)^2}} \right]
  \right)
\end{eqnarray}

\begin{figure}[ht]
 \centering
  \includegraphics[clip,width=0.65\textwidth,height=0.25\textheight]{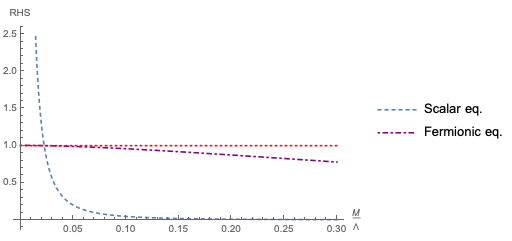} 
\caption{\it The dashed line corresponds to the right hand side of \eqref{scalynh} using $\lambda^2=0.02$ and $\bar{\mathcal{B}}=0.003$. The dot dashed line represents the right hand side of \eqref{fermarrynh} with the same values and $\bar{g}=12.6$. The dotted line represents the left hand side of the equations \eqref{scalynh} and \eqref{fermarrynh}, fixed at 1. The intersection shows there is dynamical mass generation.}\label{fig:plotbackferynh}
\end{figure}

Inserting in \eqref{fermarrynh} the value for $M$ obtained in \eqref{scalynh}, we can get the value of the four fermion interaction coupling $g/f_b$ for which we have dynamical mass generation. In figure \ref{fig:plotbackferynh} we show one example with $\bar{\mathcal{B}}=0.003$ and $\lambda^2=0.02$. The dashed line, corresponding to the right hand side Eq.~\eqref{scalynh}, gives a value of $M/\Lambda\simeq0.023$ for which \eqref{scalynh} is satisfied. Using this value in \eqref{fermarrynh} we observe that there is a consistent solution, provided that the four-fermion coupling assumes the value
\be\label{gbarval}
\bar{g} \equiv \frac{g\,\Lambda}{f_b} = 12.6\, \quad \Rightarrow \quad \frac{f_b}{g} = \frac{\Lambda}{12.6}. 
\ee
which is very close to the value \eqref{4fcL} (upon the correspondence $f_4=f_b/g$) in the absence of the axial background. 
The dotted-dashed line in fig.~\ref{fig:plotbackferynh} represents the right hand side of \eqref{fermarrynh} using these values for $\bar{\mathcal{B}}$, $\lambda$ and $\bar{g}$. We observe that the dashed and dotted-dashed lines intersect the dotted line, corresponding to the fixed value 1, at the same point. This implies that there is a consistent solution for the equations \eqref{scalynh} and \eqref{fermarrynh}, that is, there is dynamical mass generation for this case, for the  fixed value \eqref{gbarval} of the four-fermion interaction coupling.

\section{Antihermitian Background in Hermitian Yukawa Interactions- no anomalies \label{sec:nhabhy}}

We now consider the case when the axial background $\mathcal B_\mu$ is antihermitian~\cite{klevansky} and the Yukawa coupling is hermitian. The results in this case are obtained by replacing $|\mathcal B| \to i|\mathcal B|$ in \eqref{sdsolbackscal} and \eqref{sdsolbackferm}. The pertinent SD equations become

\begin{eqnarray}
\label{sdsolnhbackscal}
  M^2 &=& M_0^2 + \frac{\lambda^2}{16 \pi^2 \mathcal{B}^2}
  \left( \Lambda^2 \left( \Lambda^2 + 2 \alpha \beta - \sqrt{(\alpha^2 + \Lambda^2) (\beta^2 +
  \Lambda^2)} \right) + \frac{1}{2} \left( \sqrt{(\alpha^2 + \Lambda^2) (\beta^2 +
  \Lambda^2)} - | \alpha | \beta \right) (3 (\alpha^2 + \beta^2) - 8 \alpha \beta) \right.\nonumber\\
  & & \left. - \frac{1}{2} (\alpha -
  \beta)^2 (3 (\alpha^2 + \beta^2) - 2 \alpha \beta) \ln \left[ \frac{\sqrt{(\alpha^2 + \Lambda^2)} +
  \sqrt{(\beta^2 + \Lambda^2)}}{| \alpha | + \beta} \right] \right)
\end{eqnarray}

\begin{eqnarray}
\label{sdsolnhbackferm}
  1 &=& -\frac{\lambda^2}{32 \pi^2 \mathcal{B}^2} \left(
  \Lambda^2 - (M^2 - \alpha \beta) \left( 1 + \frac{M^2 - \alpha \beta}{\sqrt{(\alpha^2 - M^2) (\beta^2 -
  M^2)}} \right) \ln \left( 1 + \frac{\Lambda^2}{M^2} \right) - \left(
  \sqrt{(\alpha^2 + \Lambda^2) (\beta^2 + \Lambda^2)} - | \alpha | \beta \right) \right.\nonumber\\
  & & \left. + ((\alpha - \beta)^2 +
  2 (M^2 - \alpha \beta)) \ln \left[ \frac{\sqrt{(\alpha^2 + \Lambda^2)} + \sqrt{(\beta^2 +
  \Lambda^2)}}{| \alpha | + \beta} \right] \right.\nonumber\\
  & & \left. + \frac{(M^2 - \alpha \beta)^2}{\sqrt{(\alpha^2 - M^2)
  (\beta^2 - M^2)}} \ln \left[ \frac{\left( \sqrt{(\alpha^2 - M^2) (\beta^2 - M^2)} +
  \sqrt{(\alpha^2 + \Lambda^2) (\beta^2 + \Lambda^2)} \right)^2 - (M^2 +
  \Lambda^2)^2}{\left( | \alpha | \beta + \sqrt{(\alpha^2 - M^2) (\beta^2 - M^2)} \right)^2 -
  M^4} \right] \right) \nonumber\\
  & & - \frac{g^2}{64 \pi^2 f_b^2 \mathcal{B}^2} \left( \Lambda^2
  \left( \Lambda^2 + 2 \alpha \beta - \sqrt{(\alpha^2 + \Lambda^2) (\beta^2 + \Lambda^2)}
  \right) + \frac{1}{2} \left( \sqrt{(\alpha^2 + \Lambda^2) (\beta^2 + \Lambda^2)} - |
  \alpha | \beta \right) (3 (\alpha^2 + \beta^2) - 8 \alpha \beta) \right.\nonumber\\
  & & \left. - \frac{1}{2} (\alpha - \beta)^2 (3 (\alpha^2 + \beta^2)
  - 2 \alpha \beta) \ln \left[ \frac{\sqrt{(\alpha^2 + \Lambda^2)} + \sqrt{(\beta^2 +
  \Lambda^2)}}{| \alpha | + \beta} \right] \right)\, ,
\end{eqnarray}
where here $\alpha = m - i | \mathcal{B} |$ and $\beta = m + i |\mathcal{B} |$. 

\subsection{The limit $\lambda \to 0$, $g \ne 0$ \label{sec:l=0gne0}} 

We notice that in the limit $\lambda^2\to 0$, $g \ne 0$, where only the (attractive) four-fermion interaction is present, we recover the result of \cite{klevansky}. In fact, in this limit, Eqs.~\eqref{sdsolnhbackscal} and \eqref{sdsolnhbackferm} read
\begin{equation}
\label{nhbackscal}
M^2 = M^2_0.
\end{equation}
and
\begin{eqnarray}
\label{nhbackferm}
1 &=& - \frac{g^2}{64 \pi^2 f^2_b \mathcal{B}^2} \left( \Lambda^4 + \Lambda^2
\left( 2 (m^2 +\mathcal{B}^2) - \sqrt{(\Lambda^2 + m^2 -\mathcal{B}^2)^2 + 4
m^2 \mathcal{B}^2} \right) \right.\nonumber\\
& & \left. - (m^2 + 7\mathcal{B}^2) \left( \sqrt{(\Lambda^2 +
m^2 -\mathcal{B}^2)^2 + 4 m^2 \mathcal{B}^2} - (m^2 +\mathcal{B}^2) \right) \right.\nonumber\\
& & \left. + 4\mathcal{B}^2 (m^2 - 2\mathcal{B}^2) \log \left[ \frac{(m^2 -\mathcal{B}^2) +
\Lambda^2 + \sqrt{(\Lambda^2 + m^2 -\mathcal{B}^2)^2 + 4 m^2 \mathcal{B}^2}}{2
m^2} \right] \right),
\end{eqnarray}

The scalar mass is only the bare mass because, since there is no interaction in this limit. The fermionic equation \eqref{nhbackferm} is essentially the same  as equation (34) in ref.~\cite{klevansky}, where the Nambu Jona Lasinio  four-fermion model was studied in the presence of a constant  non-hermitian axial background. The conclusion in \cite{klevansky} points to the fact that inclusion of this background increases the dynamical fermion mass. In this case, of course, the condition $m \simeq M$ is not valid, in view of \eqref{nhbackscal}.

\subsection{On the (pseudo)scalar mass equation when $\lambda \ne 0$.}

Let us now study the general case, when $\lambda \ne 0$. Proceeding as in the hermitian background case, we rearrange \eqref{sdsolnhbackscal}, setting $m \simeq M$. Upon redefining $\bar{M}=M/\Lambda$,  $\bar{M}_0=M_0/\Lambda$ and $\bar{\mathcal{B}}=\mathcal{B}/\Lambda$, we have:

\begin{eqnarray}
\label{scalarnhbck}
1 &=& \frac{\bar{M}_0}{\bar{M}}+\frac{\lambda^2}{16 \pi^2 \bar{M}^2 \bar{\mathcal{B}}^2}
\left( 1 + 2 (\bar{M}^2 + \bar{\mathcal{B}}^2) - \sqrt{(1 + \bar{M}^2 -
\bar{\mathcal{B}}^2)^2 + 4 \bar{M}^2 \bar{\mathcal{B}}^2} \right.\nonumber\\
& & \left. - \left( \sqrt{(1 +
\bar{M}^2 - \bar{\mathcal{B}}^2)^2 + 4 \bar{M}^2 \bar{\mathcal{B}}^2} -
(\bar{M}^2 + \bar{\mathcal{B}}^2) \right) (\bar{M}^2 + 7 \bar{\mathcal{B}}^2)\right.\nonumber\\
& &\left. + 4 \bar{\mathcal{B}}^2 (\bar{M}^2 - 2 \bar{\mathcal{B}}^2) \log \left[
\frac{1 + \bar{M}^2 - \bar{\mathcal{B}}^2 + \sqrt{(1 + \bar{M}^2 -
\bar{\mathcal{B}}^2)^2 + 4 \bar{M}^2 \bar{\mathcal{B}}^2}}{2 \bar{M}^2}
\right] \right).
\end{eqnarray}

\begin{figure}[ht]
 \centering
  \includegraphics[clip,width=0.65\textwidth,height=0.25\textheight]{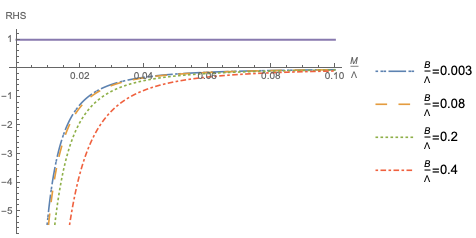} 
\caption{\it The Plot shows different values for the right hand side of \eqref{scalarnhbck} considering $\lambda^2=0.02$.  The curves do not intersect with the constant curve fixed at 1. This means that, in the absence of a bare (pseudo)scalar mass,  there is no dynamical mass for the (pseudo) scalar field, when the axial background is antihermitian.}\label{fig:plotback1b}
\end{figure}

Fig. \ref{fig:plotback1b} shows the curves corresponding to the right hand side of \eqref{scalarnhbck} with $M_0=0$ for different values of $\bar{\mathcal{B}}$. We notice that there is no intersection between these curves and the left hand side of \eqref{scalarnhbck} of value $1$, which implies that \eqref{scalarnhbck} has no solution if $M_0=0$.

On the other hand, for non-zero bare scalar mass $M_0\neq 0$, one can obtain consistent mass generation, similarly to 
the case of hermitian background. Indeed, 
in the small $|\mathcal{B}| \ll M \ll \Lambda $ limit, the value of $M_0$,  can be obtained by expanding \eqref{sdsolnhbackscal} in powers of the background, and using $m\simeq M \ll \Lambda$:

\begin{eqnarray}
\label{seriemassnh}
M_0^2 = M^2 + \frac{\lambda^2}{4 \pi^2} \Lambda^2 -
\frac{\lambda^2 \mathcal{B}^2}{8 \pi^2} \left[
7- 4 \ln \left(\frac{\Lambda^2}{M^2} \right) \right],
\end{eqnarray}
where $M$ can be found from the fermionic equation, which we now proceed to study.

\subsection{Analysis of the fermionic equation}

On considering $m \simeq M$ and using $\bar{M}=M/\Lambda$ with $\bar{\mathcal{B}}=\mathcal{B}/\Lambda$ we obtain from \eqref{sdsolnhbackferm}:

\begin{eqnarray}
\label{fermarrnhbck}
  1 &=& -\frac{\lambda^2}{32 \pi^2 \bar{\mathcal{B}}^2} \left( 1
  + \bar{\mathcal{B}}^2 \left( 1 -
  \frac{\bar{\mathcal{B}}^2}{\sqrt{\bar{\mathcal{B}}^4 + 4 \bar{M}^2
  \bar{\mathcal{B}}^2}} \right) \log \left( 1 + \frac{1}{\bar{M}^2} \right) -
  \left( \sqrt{(1 + \bar{M}^2 - \bar{\mathcal{B}}^2)^2 + 4 \bar{M}^2
  \bar{\mathcal{B}}^2} - (\bar{M}^2 + \bar{\mathcal{B}}^2) \right) \right.\nonumber\\
  & & \left. - 3 \bar{\mathcal{B}}^2 \log \left[ \frac{1 + (\bar{M}^2 -
  \bar{\mathcal{B}}^2) + \sqrt{(1 + \bar{M}^2 - \bar{\mathcal{B}}^2)^2 + 4
  \bar{M}^2 \bar{\mathcal{B}}^2}}{2\bar{M}^2} \right] \right.\nonumber\\ 
  & & \left. + \frac{\bar{\mathcal{B}}^4}{\sqrt{\bar{\mathcal{B}}^4 + 4 \bar{M}^2
  \bar{\mathcal{B}}^2}} \log \left[ \frac{\left( \sqrt{\bar{\mathcal{B}}^4 + 4
  \bar{M}^2 \bar{\mathcal{B}}^2} + \sqrt{(1 + \bar{M}^2 -
  \bar{\mathcal{B}}^2)^2 + 4 \bar{M}^2 \bar{\mathcal{B}}^2} \right)^2 - (1 +
  \bar{M}^2)^2}{\left( (\bar{M}^2 + \bar{\mathcal{B}}^2) +
  \sqrt{\bar{\mathcal{B}}^4 + 4 \bar{M}^2 \bar{\mathcal{B}}^2} \right)^2 -
  \bar{M}^4} \right] \right) \nonumber\\
  & & - \frac{\bar{g}^2}{64 \pi^2
  \bar{\mathcal{B}}^2} \left( 1 + 2 (\bar{M}^2 + \bar{\mathcal{B}}^2) -
  \sqrt{(1 + \bar{M}^2 - \bar{\mathcal{B}}^2)^2 + 4 \bar{M}^2
  \bar{\mathcal{B}}^2} \right.\nonumber\\
  & & \left. - \left( \sqrt{(1 + \bar{M}^2 - \bar{\mathcal{B}}^2)^2
  + 4 \bar{M}^2 \bar{\mathcal{B}}^2} - (\bar{M}^2 +
  \bar{\mathcal{B}}^2) \right) (\bar{M}^2 + 7 \bar{\mathcal{B}}^2) \right.\nonumber\\
  & & \left. + 4 \bar{\mathcal{B}}^2 (\bar{M}^2 - 2 \bar{\mathcal{B}}^2) \log \left[ \frac{1
  + (\bar{M}^2 - \bar{\mathcal{B}}^2) + \sqrt{(1 + \bar{M}^2 -
  \bar{\mathcal{B}}^2)^2 + 4 \bar{M}^2 \bar{\mathcal{B}}^2}}{2\bar{M}^2} \right]
  \right)
\end{eqnarray}

\begin{figure}[ht]
 \centering
  \includegraphics[clip,width=0.65\textwidth,height=0.25\textheight]{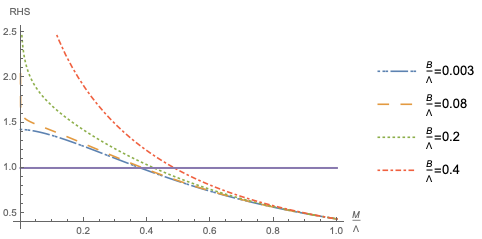} 
\caption{\it The plot shows the right hand side of \eqref{fermarrnhbck} for different possibilities of $\lambda$ and $\mathcal{B}$. All curves are computed using $\lambda^2=0.05$ and $\bar{g}=15$. The curves cross the solid line corresponding to the value $1$, which represents the left-hand side of \eqref{fermarrnhbck}. This means that there is dynamical mass generation for the fermion fields in this case.}\label{fig:plotbacknhf}
\end{figure}

In figure \ref{fig:plotbacknhf} we plot the right hand side of \eqref{fermarrnhbck} for specific values of $\lambda$ and $g$. We observe that,  for different values of $\mathcal{B}$, the various curves always intersect the value $1$, corresponding to the left hand side of \eqref{fermarrnhbck}. This implies  
the existence of a dynamical mass for fermion in the case of hermitian Yukawa interactions in the presence of antihermitian axial constant backgrounds. Upon determining $m \simeq M$, one then obtains from \eqref{seriemassnh} the required value of $M_0$.  As an example,  in table \ref{tab:m0values}, we show the values for $M$ and $M_0$ obtained from \eqref{fermarrnhbck} and \eqref{seriemassnh}, for fixed values of $\lambda^2=0.05$ and $\bar{g}=15$, and different values of $\mathcal{B}$, as in fig.~\ref{fig:plotbacknhf} . We notice that the pertinent values of $M_0$ are  positive, which proves the consistency of the approach.

\begin{table}[ht]
\setlength{\extrarowheight}{10pt}
\setlength{\tabcolsep}{12pt}
\begin{tabular}{|l|l|l|}
\hline
$\bar{\mathcal{B}}$ & $\bar{M}$ & $\bar{M}_0$ \\ \hline
$0.003$      &    $0.378$        &        $0.379$     \\ \hline
$0.08$       &      $0.388$      &      $0.389$       \\ \hline
$0.2$        &       $0.422$     &      $0.423$       \\ \hline
$0.4$        &        $0.487$    &      $0.488$       \\ \hline
\end{tabular}
\caption{Values of $M$ found from Eq.~\eqref{fermarrnhbck} and the corresponding values of $M_0$ obtained from \eqref{seriemassnh}, for the fixed values $\lambda^2=0.05$, $\bar{g}=15$ and different values  of $\bar{\mathcal{B}}$ (first column). We observe that the system of equations yields positive values for $M_0$, which implies consistent dynamical mass generation in this case.}
\label{tab:m0values}
\end{table}

The above results imply that irrespective of whether the axial background is hermitian or antihermitian, 
in the case of Hermitian Yukawa interactions, one needs to give a bare mass to the (pseudo) scalar field in order to obtain non-trivial dynamical mass generation for both the fermion and the scalar fields. As we observe from fig.~\ref{fig:plotbacknhf}, for the case of anti-hermitian axial backgrounds and hermitian Yukawa interactions, the larger the background, the bigger the mass, so the presence of the antihermitian background assists dynamical mass generation, in similar spirit to the results of \cite{klevansky} for the NJL model in the absence of Yukawa interactions.

\section{Antihermitian Background in antihermitian Yukawa Interactions- no anomalies \label{sec:nhabnhy}}

For the case of both antihermitian background and Yukawa interactions, one may obtain the pertinent solutions for mass generation by substituing $|\mathcal B| \to i|\mathcal B|$ in \eqref{sdsolbackynhscal} and \eqref{sdsolbackynhferm}. This gives

\begin{eqnarray}
\label{sdnhbnhyscal}
  M^2 &=& -\frac{\lambda^2}{16 \pi^2 \mathcal{B}^2}
  \left( \Lambda^2 \left( \Lambda^2 + 2 \alpha \beta - \sqrt{(\alpha^2 + \Lambda^2) (\beta^2 +
  \Lambda^2)} \right) + \frac{1}{2} \left( \sqrt{(\alpha^2 + \Lambda^2) (\beta^2 +
  \Lambda^2)} - | \alpha | \beta \right) (3 (\alpha^2 + \beta^2) - 8 \alpha \beta) \right.\nonumber\\
  & & \left. - \frac{1}{2} (\alpha -
  \beta)^2 (3 (\alpha^2 + \beta^2) - 2 \alpha \beta) \ln \left[ \frac{\sqrt{(\alpha^2 + \Lambda^2)} +
  \sqrt{(\beta^2 + \Lambda^2)}}{| \alpha | + \beta} \right] \right)
\end{eqnarray}
and

\begin{eqnarray}
\label{sdnhbnhyferm}
  1 &=& \frac{\lambda^2}{32 \pi^2 \mathcal{B}^2} \left(
  \Lambda^2 - (M^2 - \alpha \beta) \left( 1 + \frac{M^2 - \alpha \beta}{\sqrt{(\alpha^2 - M^2) (\beta^2 -
  M^2)}} \right) \ln \left( 1 + \frac{\Lambda^2}{M^2} \right) - \left(
  \sqrt{(\alpha^2 + \Lambda^2) (\beta^2 + \Lambda^2)} - | \alpha | \beta \right) \right.\nonumber\\
  & & \left. + ((\alpha - \beta)^2 +
  2 (M^2 - \alpha \beta)) \ln \left[ \frac{\sqrt{(\alpha^2 + \Lambda^2)} + \sqrt{(\beta^2 +
  \Lambda^2)}}{| \alpha | + \beta} \right] \right.\nonumber\\
  & & \left. + \frac{(M^2 - \alpha \beta)^2}{\sqrt{(\alpha^2 - M^2)
  (\beta^2 - M^2)}} \ln \left[ \frac{\left( \sqrt{(\alpha^2 - M^2) (\beta^2 - M^2)} +
  \sqrt{(\alpha^2 + \Lambda^2) (\beta^2 + \Lambda^2)} \right)^2 - (M^2 +
  \Lambda^2)^2}{\left( | \alpha | \beta + \sqrt{(\alpha^2 - M^2) (\beta^2 - M^2)} \right)^2 -
  M^4} \right] \right) \nonumber\\
  & & - \frac{g^2}{64 \pi^2 f_b^2 \mathcal{B}^2} \left( \Lambda^2
  \left( \Lambda^2 + 2 \alpha \beta - \sqrt{(\alpha^2 + \Lambda^2) (\beta^2 + \Lambda^2)}
  \right) + \frac{1}{2} \left( \sqrt{(\alpha^2 + \Lambda^2) (\beta^2 + \Lambda^2)} - |
  \alpha | \beta \right) (3 (\alpha^2 + \beta^2) - 8 \alpha \beta) \right.\nonumber\\
  & & \left. - \frac{1}{2} (\alpha - \beta)^2 (3 (\alpha^2 + \beta^2)
  - 2 \alpha \beta) \ln \left[ \frac{\sqrt{(\alpha^2 + \Lambda^2)} + \sqrt{(\beta^2 +
  \Lambda^2)}}{| \alpha | + \beta} \right] \right)\,,
\end{eqnarray}
where here $\alpha = m - i| \mathcal{B} |$ and $\beta = m + i|\mathcal{B} |$.

\subsection{Dynamical (pseudo)scalar mass}

As before, for solutions $m\simeq M$ and redefining $\bar{M}=M/\Lambda$ and $\bar{\mathcal{B}}=\mathcal{B}/\Lambda$ we get

\begin{eqnarray}
\label{scalnhbnhy}
1 &=& -\frac{\lambda^2}{16 \pi^2 \bar{M}^2 \bar{\mathcal{B}}^2}
\left( 1 + 2 (\bar{M}^2 + \bar{\mathcal{B}}^2) - \sqrt{(1 + \bar{M}^2 -
\bar{\mathcal{B}}^2)^2 + 4 \bar{M}^2 \bar{\mathcal{B}}^2} \right.\nonumber\\
& & \left. - \left( \sqrt{(1 +
\bar{M}^2 - \bar{\mathcal{B}}^2)^2 + 4 \bar{M}^2 \bar{\mathcal{B}}^2} -
(\bar{M}^2 + \bar{\mathcal{B}}^2) \right) (\bar{M}^2 + 7 \bar{\mathcal{B}}^2)\right.\nonumber\\
& &\left. + 4 \bar{\mathcal{B}}^2 (\bar{M}^2 - 2 \bar{\mathcal{B}}^2) \log \left[
\frac{1 + \bar{M}^2 - \bar{\mathcal{B}}^2 + \sqrt{(1 + \bar{M}^2 -
\bar{\mathcal{B}}^2)^2 + 4 \bar{M}^2 \bar{\mathcal{B}}^2}}{2 \bar{M}^2}
\right] \right).
\end{eqnarray}

\begin{figure}[ht]
 \centering
  \includegraphics[clip,width=0.65\textwidth,height=0.25\textheight]{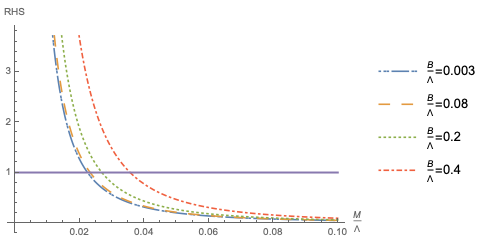} 
\caption{\it The Plot shows different values of the right hand side of \eqref{scalnhbnhy} for the fixed value $\lambda^2=0.02$.  
All the curves intersect the solid line, which represents the left hand side of \eqref{scalnhbnhy}. It means that, for non-Hermitian Yukawa coupling and non-hermitian background, there is dynamical mass generation for the scalar, without the need to introduce a bare mass.}\label{fig:plotscalnhbnhy}
\end{figure}

The situation is similar to the non-hermitian Yukawa coupling and hermitian background, for which one does not need a bare mass for the scalar to generate dynamical mass. Figure \ref{fig:plotscalnhbnhy} shows the plot of the right hand side of \eqref{scalnhbnhy} as a function of $\bar M$ for different values of $\mathcal{B}$. All the curves intersect the solid line, corresponding to the fixed value 1, which represents the left hand side of \eqref{scalnhbnhy}. 

Comparing figures \ref{fig:plotbacksynh} and \ref{fig:plotscalnhbnhy}, where in both cases the Yukawa interaction is non-hermitian, we observe that, for a non-hermitian axial background, the larger the background, the bigger the dynamical mass. This is to be contrasted with the case of a hermitian axial  background, for which, as we have seen in section \ref{sec:hbsm}, the larger the background, the smaller the dynamical mass.

\subsection{Fermionic equation}

Proceeding as before, for $m\simeq M$ with the redefinitions $\bar{M}=M/\Lambda$, $\bar{\mathcal{B}}=\mathcal{B}/\Lambda$ and $\bar{g}=(g\Lambda)/f_b$ the equation \eqref{sdnhbnhyferm} becomes:

\begin{eqnarray}
\label{fermnhbnhy}
  1 &=& \frac{\lambda^2}{32 \pi^2 \bar{\mathcal{B}}^2} \left( 1
  + \bar{\mathcal{B}}^2 \left( 1 -
  \frac{\bar{\mathcal{B}}^2}{\sqrt{\bar{\mathcal{B}}^4 + 4 \bar{M}^2
  \bar{\mathcal{B}}^2}} \right) \log \left( 1 + \frac{1}{\bar{M}^2} \right) -
  \left( \sqrt{(1 + \bar{M}^2 - \bar{\mathcal{B}}^2)^2 + 4 \bar{M}^2
  \bar{\mathcal{B}}^2} - (\bar{M}^2 + \bar{\mathcal{B}}^2) \right) \right.\nonumber\\
  & & \left. - 3 \bar{\mathcal{B}}^2 \log \left[ \frac{1 + (\bar{M}^2 -
  \bar{\mathcal{B}}^2) + \sqrt{(1 + \bar{M}^2 - \bar{\mathcal{B}}^2)^2 + 4
  \bar{M}^2 \bar{\mathcal{B}}^2}}{2\bar{M}^2} \right] \right.\nonumber\\ 
  & & \left. + \frac{\bar{\mathcal{B}}^4}{\sqrt{\bar{\mathcal{B}}^4 + 4 \bar{M}^2
  \bar{\mathcal{B}}^2}} \log \left[ \frac{\left( \sqrt{\bar{\mathcal{B}}^4 + 4
  \bar{M}^2 \bar{\mathcal{B}}^2} + \sqrt{(1 + \bar{M}^2 -
  \bar{\mathcal{B}}^2)^2 + 4 \bar{M}^2 \bar{\mathcal{B}}^2} \right)^2 - (1 +
  \bar{M}^2)^2}{\left( (\bar{M}^2 + \bar{\mathcal{B}}^2) +
  \sqrt{\bar{\mathcal{B}}^4 + 4 \bar{M}^2 \bar{\mathcal{B}}^2} \right)^2 -
  \bar{M}^4} \right] \right) \nonumber\\
  & & - \frac{\bar{g}^2}{64 \pi^2
  \bar{\mathcal{B}}^2} \left( 1 + 2 (\bar{M}^2 + \bar{\mathcal{B}}^2) -
  \sqrt{(1 + \bar{M}^2 - \bar{\mathcal{B}}^2)^2 + 4 \bar{M}^2
  \bar{\mathcal{B}}^2} \right.\nonumber\\
  & & \left. - \left( \sqrt{(1 + \bar{M}^2 - \bar{\mathcal{B}}^2)^2
  + 4 \bar{M}^2 \bar{\mathcal{B}}^2} - (\bar{M}^2 +
  \bar{\mathcal{B}}^2) \right) (\bar{M}^2 + 7 \bar{\mathcal{B}}^2) \right.\nonumber\\
  & & \left. + 4 \bar{\mathcal{B}}^2 (\bar{M}^2 - 2 \bar{\mathcal{B}}^2) \log \left[ \frac{1
  + (\bar{M}^2 - \bar{\mathcal{B}}^2) + \sqrt{(1 + \bar{M}^2 -
  \bar{\mathcal{B}}^2)^2 + 4 \bar{M}^2 \bar{\mathcal{B}}^2}}{2\bar{M}^2} \right]
  \right)
\end{eqnarray}

\begin{figure}[ht]
 \centering
  \includegraphics[clip,width=0.65\textwidth,height=0.25\textheight]{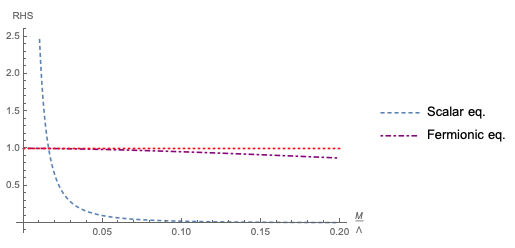} 
\caption{\it The dashed line corresponds to the right hand side of \eqref{scalnhbnhy} for $\lambda^2=0.01$ and $\bar{\mathcal{B}}=0.0004$. The dotted-dashed line represents the right hand side of \eqref{fermnhbnhy} for the above values of $\lambda^2$ and $\bar{\mathcal{B}}$, and $\bar{g}=12.58$. The constant dotted line at 1 represents the left hand side of the equations \eqref{scalnhbnhy} and \eqref{fermnhbnhy}. The existence of a common intersection point for all three  curves demonstrates the existence  of dynamical mass generation.}\label{fig:plotnhbackferynh}
\end{figure}

From \eqref{scalnhbnhy} one obtains a value for $M$, and then, upon inserting it in \eqref{fermnhbnhy}, one determines the value of the four-fermion coupling $g/f_b$ for which there is a consistent dynamical mass, as in the hermitian background case.  As a concrete example, in fig.~\ref{fig:plotnhbackferynh} we plot  the right hand side of Eqs.~\eqref{scalnhbnhy} and \eqref{fermnhbnhy}, for $\bar{\mathcal{B}}=0.0004$ and $\lambda^2=0.01$. We observe from the figure that the dashed line, corresponding to the right hand side of \eqref{scalnhbnhy}, intersects the constant dotted line at 1, representing the left hand side of the equation, which implies the existence of a non-trivial solution for $M/\Lambda=0.159$. Using this value in \eqref{fermnhbnhy}, we then obtain a consistent solution for $\bar{g}=12.58$, in the sense that for this value of $\bar{g}$ the three curves have a common intersection at $M/\Lambda=0.159$.
This demonstrates that there is dynamical mass generation with $m \simeq M < \Lambda$ in this case.\footnote{In this example, the mass is an order of magnitude smaller than the cutoff, which is in agreement with our approximation $M^2 \ll \Lambda^2$, since the square of the mass will be two orders of magnitude smaller than the square of the cutoff, thus justifying the analytic approximations made in our work. However, we note that there are other examples, not mentioned explicitly, in which $m \simeq M \ll \Lambda$.}

\section{Hermitian Anomaly term and Hermitian Yukawa Interaction - no Background \label{sec:hahy}}

In this and the next sections we shall consider the effects of the anomaly terms in \eqref{bacanomlag}
($\gamma \ne 0$) in the absence of a constant axial background. This is the situation encountered in the model of \cite{mp} ({\it cf.} \eqref{bacoupl3}, \eqref{anomfermion}). 
The Lagrangian for this case reads 
\begin{equation}\label{gamg}
  \mathcal{L}= \frac{1}{2} \partial_{\mu} a \partial^{\mu} a + \bar{\psi} i \slashed{\partial} \psi - \frac{\gamma}{f_b}
  (\partial_{\mu} a) \bar{\psi} \gamma^{\mu} \gamma^5 \psi + i
  \lambda a \bar{\psi} \gamma^5 \psi - \frac{g^2}{2 f^2_b}
  (\bar{\psi} \gamma^5 \psi)^2
\end{equation}
Motivated by the physics of the model \cite{mp}, in which the anomaly term coefficient is related to a shift-symmetry-respecting kinetic mixing between the gravitational and standard axion fields, while the 
Yukawa interactions are associated with shift-symmetry breaking non-perturbative (e.g. string instanton) effects, it is natural to assume the following regime of the various couplings appearing in \eqref{gamg}:
\be\label{rangegamg}
 |g| \, \gtrsim 1 \, \gg \, |\gamma | \, > \,  |\lambda | \, > 0,
\ee
which we restrict our attention to in what follows (except when we discuss the $g =0$ limit, in which only 
$1 \gg |\gamma| > |\lambda| > 0$ will be assumed).

After the standard linearization of the four fermion interactions by means of the auxiliary  scalar field $\sigma$, the generating functional,  is given by

\begin{eqnarray}
  Z [K, J, \eta, \bar{\eta}] &=& \int \mathcal{D} [\sigma a \psi \bar{\psi}]
  \exp \left\{ i \int d^4 x \left[ \frac{1}{2} \partial_{\mu} a \partial^{\mu}
  a + \bar{\psi} i \slashed{\partial} \psi - \frac{\gamma}{f_b}
  (\partial_{\mu} a) \bar{\psi} \gamma^{\mu} \gamma^5 \psi + i
 \lambda a \bar{\psi} \gamma^5 \psi - \frac{f^2_b}{2 g^2} \sigma^2 -
  i \sigma \bar{\psi} \gamma^5 \psi \right] \right.\nonumber\\
  & & \left. + i \int d^4 x [J a + \bar{\psi} \eta + \bar{\eta} \psi + K \sigma] \right\}
\end{eqnarray}

The inclusion of the anomaly term will modify the vertex $a\bar{\psi}\psi$ with a term proportional to the scalar field momentum. Therefore, the Feynman rule for this (bare) vertex becomes: $-\left(\lambda + \frac{\gamma}{f_b} p_{\mu}
\gamma^{\mu}\right)\gamma^5$, with the convention that all the momenta are pointing towards the vertex, and $p_{\mu}$ is the momentum of the scalar field. The rule for the (bare) vertex $\sigma\bar{\psi}\psi$ is $\gamma^5$. 

The corresponding SD equations  then read:
\begin{equation}
  G^{- 1}_f (k) - S^{- 1}_f (k) = \left( \int_p  \left( \lambda -
  \frac{\gamma}{f_b} \left( \slashed{p} - \slashed{k} \right) \right)
  \gamma^5 G_f (p) \Gamma^{(3)} (p, k) G_s (p - k) \right) - \gamma^5 \left(
  \int  \frac{d^4 p}{(2 \pi)^4} G_f (p) \Gamma_2^{(3)} (p, k) G_{\sigma} (p -
  k) \right)
\end{equation}
\begin{equation}
  G^{- 1}_s (k) - S^{- 1}_s (k) = - \tmop{tr} \left[ \int_p  \left(
 \lambda + \frac{\gamma}{f_b} \slashed{k} \right) \gamma^5
  G_f (p) \Gamma^{(3)} (p, k) G_f (p - k) \right]\,,
\end{equation} 
where  the dressed fermion ($G_f$), scalar $(G_s$) and auxiliary scalar ($G_\sigma$) 
propagators 
are given by~\cite{amsot}:\\
$G_f (p) = \frac{i}{\slashed{p} - m}$, $G_s=\frac{i}{p^2-M^2}$, and $G_\sigma (p) = -\frac{i\, g^2}{f_b^2}$, where $M$ and $m$ denote the dynamical scalar and Dirac-fermion masses,  respectively (as already mentioned, here and in \cite{amsot}, we set the axial fermion mass to zero, $\mu=0$). The quantities $\Gamma^{(3)} (p, k)$ and $\Gamma_2^{(3)} (p, k)$ denote the appropriate dressed vertices 
for the scalar-fermion and auxiliary-scalar-fermion trilinear interactions. As in \cite{amsot}, we shall be using the rainbow approximation, where quantum corrections to the vertices are ignored. This suffices for the case of small $|\gamma|, |\lambda| \ll 1$ we consider here.

 In the absence of external momenta, then,  the SD equations become:

\begin{equation}
\label{sdh1}
  G^{- 1}_f (0) - S^{- 1}_f (0) = - \left( \int_p \left( \lambda - 
  \frac{\gamma}{f_b} \slashed{p} \right) \gamma^5 G_f (p) \left(
  \lambda - \frac{\gamma}{f_b} \slashed{p} \right) \gamma^5
  G_s (p) \right) - \gamma^5 \left( \int_p G_f (p) \gamma^5 G_{\sigma} (p)
  \right)
\end{equation}
\begin{equation}
\label{sdh2}
  G^{- 1}_s (0) - S^{- 1}_s (0) = \lambda^2 \tmop{tr} \left[
  \int_p \gamma^5 G_f (p) \gamma^5 G_f (p) \right]
\end{equation}
The reader should notice that the scalar SD equation (\ref{sdh2}) is independent of the anomaly coefficient $\gamma$. 

As in the previous cases, we will consider a scalar bare mass $M_0$. After standard manipulations~\cite{amsot} the equations become:

\begin{eqnarray}
\label{sol1herm}
  1 &=& \frac{\lambda^2}{16 \pi^2} \frac{1}{M^2 - m^2} \left[
  M^2 \ln \left( 1 + \frac{\Lambda^2}{M^2} \right) - m^2 \ln \left( 1 +
  \frac{\Lambda^2}{m^2} \right) \right] \nonumber\\
  & & - \frac{\gamma^2}{16
  \pi^2 f^2_b} \frac{1}{(M^2 - m^2)} \left[ - \Lambda^2 (M^2 - m^2) + M^4 \ln
  \left( 1 + \frac{\Lambda^2}{M^2} \right) - m^4 \ln \left( 1 +
  \frac{\Lambda^2}{m^2} \right) \right] \nonumber\\
  & & + \frac{g^2}{16 \pi^2 f^2_b} \left(
  \Lambda^2 - m^2 \ln \left( 1 + \frac{\Lambda^2}{m^2} \right) \right)
\end{eqnarray}
and 
\begin{equation}
\label{sol2herm}
  M^2 = M_0^2 - \frac{\lambda^2}{4 \pi^2} \left[ \Lambda^2 - m^2
  \ln \left( 1 + \frac{\Lambda^2}{m^2} \right) \right]
\end{equation}

From (\ref{sol2herm}) is easy to see that there is no solution for the scalar mass if $M_0=0$. 

We seek solutions for which $m \simeq M$. The equations for this case read
\begin{align}
\label{hermmsimM1}
  1 &= \frac{\lambda^2}{16 \pi^2} \ln \left( 1 +
  \frac{\Lambda^2}{M^2} \right) - \frac{\gamma^2}{16 \pi^2
  f^2_b} \left[ 2\, M^2\, \ln \left( 1 + \frac{\Lambda^2}{M^2} \right) -
  \Lambda^2 \right] + \frac{g^2}{16 \pi^2 f^2_b} \left( \Lambda^2 - M^2 \ln
  \left( 1 + \frac{\Lambda^2}{M^2} \right) \right)   \nonumber \\
 \Rightarrow \quad 1 &=  \frac{1}{16 \pi^2} \Big(\lambda^2 - (2\gamma^2 + g^2) \, \frac{M^2}{f_b^2} \Big) \ln\left( 1 +
  \frac{\Lambda^2}{M^2} \right) + \frac{\gamma^2 + g^2}{16\pi^2 } \, \frac{\Lambda^2 }{f_b^2} 
 \end{align}
and
\begin{equation}
\label{hermmsimM2}
  \frac{4 \pi^2}{\lambda^2} (M^2_0 - M^2) = \left[ \Lambda^2
  - M^2 \ln \left( 1 + \frac{\Lambda^2}{M^2} \right) \right]
\end{equation}

Using $\Lambda^2\gg M^2$, from \eqref{hermmsimM1} and \eqref{hermmsimM2} we 
observe that a consistent solution, in the range \eqref{rangegamg} of the parameters, is:  

\begin{align}
\label{manh}
 \frac{f_b}{\sqrt{\gamma^2 + g^2}} &\simeq \frac{\Lambda}{4
\pi} \nonumber \\
  M^2 \simeq m^2 &= \lambda^2 \frac{f_b^2}{(2\gamma^2 + g^2)} \simeq \frac{\lambda^2}{16\pi^2} \, \frac{\gamma^2 + g^2}{2\gamma^2 + g^2} \, \Lambda^2 \, \ll \Lambda^2, 
 \nonumber \\
 M_0^2 &= \frac{\lambda^2}{16\pi^2} \, \Big(\frac{9\gamma^2 + 5g^2}{2\gamma^2 + g^2}\Big) \, \Lambda^2 + \mathcal O \Big(\lambda^4 \, \ln (\lambda^2) \Big). 
  \end{align}

The above result shows that one can have dynamical mass generation for the fermion and scalar fields in the presence of the bare mass $M_0$, given by the relation \eqref{manh}. In addition, we notice from \eqref{manh} that $\gamma^2$ appears in the solution with the same sign as $g^2$. Hence, the anomaly term plays a similar r\^ole to the four fermion attractive interactions, as far as dynamical mass generation is concerned. The reader can of course check that in the limit $\gamma=g=0$, 
the SD equatIons \eqref{hermmsimM1} and \eqref{hermmsimM2} reduce to the corresponding ones in \cite{amsot}, leading to the non-perturbative solution \eqref{SD1cb}.

\section{Non-Hermitian Anomaly term and Non-Hermitian Yukawa Interaction - no Background \label{sec:nhanhy}}

The situation changes when we consider both, non-hermitian Yukawa coupling and anomaly terms. For this case, the euclidean generating functional, under the linearization of the four-fermion interactions by means of the auxiliary scalar $\sigma$, reads:

\begin{eqnarray}
  Z [K, J, \eta, \bar{\eta}] &=& \int \mathcal{D} [\sigma a \psi \bar{\psi}]
  \exp \left\{ - \int d^4 x \left[ \frac{1}{2} \partial_{\mu} a \partial^{\mu}
  a + \bar{\psi} i \slashed{\partial} \psi + i \frac{\gamma}{f_b}
  (\partial_{\mu} a) \bar{\psi} \gamma^{\mu} \gamma^5 \psi - \lambda
  a \bar{\psi} \gamma^5 \psi + \frac{f^2_b}{2 g^2} \sigma^2 - i \sigma
  \bar{\psi} \gamma^5 \psi \right] \right.\nonumber\\
   & & \left. - \int d^4 x [J a + \bar{\psi} \eta +
  \bar{\eta} \psi + K \sigma] \right\}
\end{eqnarray}

The Feynman rule for the (bare) vertex $a\bar{\psi}\psi$ is given by $\left(\lambda - \frac{\gamma}{f_b} \slashed{p} \right)\gamma^5$, in the convention where all the momenta are incoming to the vertex.

Proceeding as before, using the rainbow approximation, we arrive at the following system of SD equations with vanishing external momenta, describing the dynamical generation of fermion and scalar masses

\begin{equation}
  G^{- 1}_f (0) - S^{- 1}_f (0) = - \int_p \left( \lambda +
  \frac{\gamma}{f_b} \slashed{p} \right) \gamma^5 G_f (p) \left(
 \lambda + \frac{\gamma}{f_b} \slashed{p} \right) \gamma^5
  G_s (p) + \gamma^5 \int_p G_f (p) \gamma^5 G_{\sigma} (p),
\end{equation}
\begin{equation}
  G^{- 1}_s (0) - S^{- 1}_s (0) = \lambda^2 \tmop{tr} \left[
  \int_p \gamma^5 G_f (p) \gamma^5 G_f (p) \right].
\end{equation}

Solving these equations (setting the axial fermion mass to zero,  $\mu=0$, as above and in \cite{amsot}) we obtain 

\begin{eqnarray}
\label{sol1nohermb}
  1 &=& - \frac{\lambda^2}{16 \pi^2} \frac{1}{M^2 - m^2} \left[
  M^2 \ln \left( 1 + \frac{\Lambda^2}{M^2} \right) - m^2 \ln \left( 1 +
  \frac{\Lambda^2}{m^2} \right) \right] \nonumber\\
  & & + \frac{\gamma^2}{16
  \pi^2 f^2_b} \frac{1}{M^2 - m^2} \left[ - \Lambda^2 (M^2 - m^2) + M^4 \ln
  \left( 1 + \frac{\Lambda^2}{M^2} \right) - m^4 \ln \left( 1 +
  \frac{\Lambda^2}{m^2} \right) \right] \nonumber\\
  & & + \frac{g^2}{16 \pi^2 f^2_b} \left(
  \Lambda^2 - m^2 \ln \left( 1 + \frac{\Lambda^2}{m^2} \right) \right)
\end{eqnarray}
and
\begin{equation}
\label{sol2nohermb}
  M^2 = \frac{\lambda^2}{4 \pi^2} \left[ \Lambda^2 - m^2 \ln
  \left( 1 + \frac{\Lambda^2}{m^2} \right) \right]
\end{equation}

We seek solutions for which $m \simeq M$. In this case we have 

\begin{equation}
\label{nhanf}
  1 = - \frac{1}{16 \pi^2} \left( \lambda^2 -
  \frac{\gamma^2 M^2}{f^2_b} \right) \ln \left( 1 +
  \frac{\Lambda^2}{M^2} \right) + \frac{g^2-\gamma^2}{16
  \pi^2 f^2_b} \left[ \Lambda^2 - m^2 \ln \left( 1 + \frac{\Lambda^2}{M^2}
  \right) \right]
\end{equation}
and
\begin{equation}
\label{nhans}
\frac{4 \pi^2}{\lambda^2} M^2 = \left[ \Lambda^2 - m^2 \ln
\left( 1 + \frac{\Lambda^2}{m^2} \right) \right]
\end{equation}

Considering $\Lambda^2\gg M^2$, $|\lambda|, |\gamma| \ll 1$,  with $g \ne \gamma$,  and substituting \eqref{nhans} in \eqref{nhanf}, we obtain:
\begin{equation}
\label{nhmas}
M^2\simeq m^2 \simeq \frac{4 f^2_b \lambda^2}{g^2-\gamma^2} + | \mathcal{O} \Big((\lambda^4, \lambda^2 \gamma^2)\, \ln(\lambda^2)\Big) |
\end{equation}

Hence, consistency between \eqref{nhmas} and \eqref{nhans} requires 

\begin{equation}
\frac{f_b}{\sqrt{g^2-\gamma^2}} \simeq \frac{\Lambda}{4
\pi} - | \mathcal{O} (\lambda^2) | \quad \Rightarrow \quad M^2 \simeq m^2 \simeq \frac{\lambda^2}{4\pi^2}\, \Lambda^2 +  | \mathcal{O} \Big((\lambda^4, \lambda^2 \gamma^2)\, \ln(\lambda^2)\Big) |\,.
\end{equation}

In order to have dynamical mass generation one must have $g^2>\gamma^2$, which is satisfied if we adopt the regime of parameters \eqref{rangegamg}.\footnote{The limiting case  $g^2=\gamma^2$ is not considered in detail here, as it is not motivated by the physics of \cite{mp}. However, it is worthy of mentioning that  when $g^2=\gamma^2$, we observe from \eqref{nhanf} that, for $\gamma^2 M^2/f_b^2 > \lambda^2 $ one obtains: 
$$ M^2 \simeq m^2 \simeq \exp\Big(-\frac{16\pi^2}{\gamma^2 M^2/f_b^2  - \lambda^2}\Big) \, \Lambda^2\,, $$
which, is compatible with \eqref{nhans}, provided 
$\frac{16\pi^2}{\frac{\gamma^2 \, M^2}{f_b^2} -\lambda^2} \sim - \ln(\frac{\lambda^2}{4\pi^2}) $, that is:
$$ M^2 \simeq - \frac{f_b^2}{\gamma^2} \frac{16\pi^2}{\ln(\frac{\lambda^2}{4\pi^2})} + \frac{f_b^2}{\gamma^2} \lambda^2 \simeq \frac{\lambda^2}{4\pi^2}\, \Lambda^2  + \dots \quad \Rightarrow \quad \frac{f_b^2}{\gamma^2} \simeq - \lambda^2 \frac{\Lambda^2}{64\pi^4}\, \ln(\frac{\lambda^2}{4\pi^2}) + \dots, 
 \, \quad |\lambda |\, ,\, |\gamma| \ll 1, $$
 where the $\dots$ denote subleading terms of $\mathcal O\Big(\lambda^4 \ln(\lambda^2)\Big)$, for $|\lambda | \ll 1.$ } 
Thus in this case, contrary to the hermitian one, it is a sufficiently strong attractive four fermion interaction that drives dynamical mass generation, since the non-hermitian anomaly term resists mass generation.

\section{Conclusions and Outlook \label{sec:concl}}

Let us summarize the main points of our work, where we considered separately the effects of constant axial background and anomalies on Yukawa-interaction-induced dynamical fermion and (pseudo)scalar masses.  In the absence of anomalous terms, a (small) compared to the mass, constant in magnitude, {\it anti-hermitian} axial background {\it assists} dynamical mass generation, in both the cases of  hermitian and anti-hermitian Yukawa interactions, in the sense that the larger the magnitude of the antihermitian background, the larger the dynamical mass. The situation is {\it opposite} when the axial background is {\it hermitian}. The larger the background the smaller the dynamical mass. 
Above we considered the generation of dynamical masses for the scalars and fermions of approximately equal in magnitude, induced  by the presence of small Yukawa interaction (the only exception, was in section \ref{sec:l=0gne0}, where we considered only the effects of the antihermitian background on 
the dynamical fermion mass, induced by attractive four-fermion interactions \eqref{fourf}, in the {\it absence} of Yukawa interactions, in order to compare our results with those of \cite{klevansky}). 

We have also seen above that, in case when the Yukawa interaction is hermitian, a bare mass for the (pseudo)scalar field is required in order to have dynamical mass for the scalar and fermion fields,  independently of the hermitian or anti-hermitian nature of the background and the anomaly terms. The presence of bare mass is not necessary for dynamical mass generation in the case of non-hermitian Yukawa coupling. 

In the presence of sufficiently strong {\it attractive} four fermion interactions \eqref{fourf}, 
dynamical generation of scalar and fermion masses, similar in magnitude, is possible, 
irrespective of whether 
the background is hermitian or not. 
The value for the four fermion interaction coupling $g/f_b$ can be determined by consistency, and turns out to be of the same order of magnitude as the UV cutoff. 

When anomaly terms are present, in the absence of axial backgrounds, 
a case that allows direct comparison with the model  \eqref{bacoupl3}  of \cite{mp}, which motivated our works here and in \cite{amsot}, we observe that 
the hermitian anomaly acts in a similar way as the four fermion interaction, assisting the generation of Yukawa-induced dynamical mass for fermions and scalars. On the contrary, a non-hermitian anomaly term acts as a repulsive force and a sufficiently strong four fermion interaction is required for antihermitian-Yukawa interaction-induced dynamical mass generation.

We complete our discussion by performing a comparison between the dynamical fermion mass in the presence of anomalies with the radiative sterile neutrino mass \eqref{massR}, \eqref{massR2}. 
To this end, we need first to extend the discussion to Majorana fermion case. As already mentioned, and  
studied in some detail in \cite{amsot}, this is straightforward, and the results for the dynamical Majorana fermion mass will differ from the Dirac case studied above only by numerical factors of order 2. 
Specifically, let us commence our discussion with the hermitian anomaly and Yukawa interaction terms, \eqref{manh} in the absence of four fermion interactions ($g \to 0$). 

In the Majorana case  it can be seen, by repeating the SD analysis leading to \eqref{manh} for Majorana spinors, 
$\psi^M = \psi^C + \psi$, that the anomaly-induced dynamical mass is of the form
\be\label{majanom}
 M_{\rm Dyn~Maj}^2 \simeq m_{\rm Dyn~maj}^2 = \lambda^2 \frac{f_b^2}{2\gamma^2} \simeq \frac{\lambda^2}{16\pi^2} \,\, \Lambda^2 \, \quad \Rightarrow \quad M_{\rm Dyn~Maj} \simeq m_{\rm Dyn~Maj} \simeq \lambda \frac{\Lambda}{4\pi}, \, \quad 0 < \lambda \ll 1\,,
\ee
for appropriate values of the bare scalar mass which we do not give here, as they are not relevant for our discussion. The reader should observe that, although for consistency of our SD treatment we require 
$|\gamma |, |\lambda | \ll 1 $, in the limit $g \to 0$, the dynamical mass is actually independent of the precise magnitude of the anomaly coefficient $\gamma$.

Comparing with the radiative fermion mass \eqref{massR} (for $c_1=1$ and $\gamma \ll 1$, such that $\gamma/\sqrt{1-\gamma^2} \simeq \gamma$), which we stress again is independent of the axion mass terms, we observe that, 
\begin{align}\label{massRcomp}
M_R \sim \frac{\sqrt{3/2}\,\gamma\,  (\kappa \Lambda)^8}{24576 \,\pi^3} M_{\rm Dyn~Maj} \simeq 1.6 \cdot 10^{-6} \, \gamma \, (\kappa \Lambda)^8\, M_{\rm Dyn~Maj} \, \ll \, M_{\rm Dyn~Maj}\,,  
\end{align}
for any value of a sub-planckian cutoff $\kappa \, \Lambda \,  \lesssim 1$, and kinetic mixing coefficient $|\gamma| \ll 1|$, required or consistency of our SD analysis. 
Thus,  in this case, the anomalies-induced dynamical mass will be dominant over the anomalously generated one. This is a novel and important result that affects the conclusions of \cite{mp}, in case of course there is  a bare mass for the axions of the appropriate value to allow for the appearance of a dynamical mass. If this criterion is not satisfied, then the radiative anomalous mass~\cite{mp}, \eqref{massR}, is the only mass component for the sterile neutrinos, and the conclusions of \cite{mp} remain unaltered. 

\nicktext{We also remark at this point that, since the mass scale \eqref{majanom} also charactrerises the axion dynamical mass, then, depending on
the relative magnitude of $\lambda$ with respect to the pertinent scale in ordinary axion physics, 
the masses \eqref{majanom} could be much larger than the QCD axion mass scale 
\be\label{axionmass2}
m_{\mathcal A_{\rm QCD}} = 6 \times 10^{-10} \, \Big(\frac{10^{16} \, {\rm GeV}}{f_a}\Big)\, {\rm ev}.
\ee
where $f_a$ is the axion constant (of dimensions of mass). Experimental searches exclude $f_a < {\mathcal O}(10^9)~{\rm GeV}$. 
For instance, to accommodate Majorana right-handed fermion masses of order $10^5$~GeV, as required for the CPT leptogenesis scenario of \cite{decesare}, which involves KR primordial axion backgrounds, one needs $\lambda \simeq 10^{-15}$, leading to heavy axions of mass $10^5$~GeV. The detailed cosmology of such string-inspired field theories lies outside the phenomenological scope of our article.}

The reader should also bear in mind that the above conclusion \eqref{massRcomp} holds only for our specific model considered in this work, where the pseudoscalar field has no other potential term apart from  a bare mass. Unlike the  radiative mass, the dynamically generated one depends on the form of the scalar potential. For instance, as we have already mentioned, in case one has a Yukawa interaction with a coupling that is common with a quartic interaction for the scalar~\cite{bashir}, then the 
dynamical fermion mass is generated only above a critical value for this coupling, unlike our case, where the mass-inducing Yukawa interaction coupling $\lambda$ is very small. 

In the non-hermitian anomaly and ~Yukawa interaction case, on the other hand, 
for Majorana fermions, one would obtain the analogue of \eqref{nhmas}: 
$ M^2_{\rm Dyn~Maj} \simeq \frac{2 f_b^2\, \lambda^2}{g^2-\gamma^2} + \dots$, with $\frac{f_b}{\sqrt{g^2-\gamma^2}} \simeq \frac{\Lambda}{2\sqrt{2} \pi}$, which imply:
\begin{align}\label{massR2comp}
M_{\rm Dyn~Maj}^{anti-herm} \sim  \frac{\lambda }{2 \pi} \, \Lambda, \quad |\lambda| \ll 1. 
\end{align}
The reader should observe, again, that, although the consistency of our SD analysis requires  a non-zero but small value, $|\gamma| \ll 1$, for the anomaly coefficient, the latter does not enter explicitly the final expression for the dynamical mass, when the latter is expressed in terms of the UV cutoff. Any dependence on $\gamma$ is hidden in the relation between $f_b$ and $\Lambda$. 

Upon comparison of \eqref{massR2comp} with \eqref{massR2} (ignoring the physically irrelevant overall minus sign, 
and considering again $c_1=1$, and $|\tilde \gamma| = |\gamma| \ll 1, \, |\lambda |\ll 1 $, as required or consistency of our SD treatment), one obtains:
\begin{align}\label{massR2compb}
M_R^{\rm anti-herm} \, \sim \,  \frac{\sqrt{3/2}\, \gamma\  (\kappa \, \Lambda)^8}{49152 \,
\pi^3}\, M_{\rm Dyn~Maj}^{\rm anti-herm} \simeq 8 \cdot 10^{-7} \, \gamma \, (\kappa \, \Lambda)^8 
\, M_{\rm Dyn~Maj}^{\rm anti-herm} \ll M_{\rm Dyn~Maj}^{\rm anti-herm}\; ,
\end{align}
for any value of a sub-planckian cutoff $\kappa \, \Lambda \,  \lesssim 1$, and kinetic mixing coefficient $|\gamma| \ll 1$, required or consistency of our SD analysis. Hence again, in the non-hermitian case, the anomaly induced dynamical mass dominates over the radiative anomalously-induced one. In this case, there is no bare mass for the axion, but one needs specific four-fermion interaction  couplings  to obtain a non-trivial dynamical mass. If the latter criterion is not satisfied, then the radiative mass \eqref{massR2} is the only type of mass which is generated due to the anomalies. 

We note at this point that the dynamical masses for both (pseudo)scalars and fermions discussed above
are small, induced by small Yukawa and anomaly interaction terms, indicating potential applications of this work to ultra light axion-like particles (ALP) and sterile neutrinos, that might be of relevance for the physics of dark matter. The models discussed here, involve interactions between axions and sterile neutrinos, which might characterise multi-component dark matter models, and have implications for the Physics of the early Universe. Of particular interest in this respect are the non hermitian models, which do not require bare (pseudo)scalar-field masses for the dynamical generation of mass for the fermion and (pseudo)scalar fields. It would be interesting to examine the cosmology of such models.  However, for realistic models we need more than one flavour of neutrinos, which implies that we need to extend the above considerations into theories involving fermion mixing. For non hermitian models this is a non trivial task, and it is only recently that such a task is initiated in \cite{mixing}, but for the simpler case of PT symmetric scalar field theories.

 We would like to close by stressing once more that, in our approach, when dealing with the non-hermitian cases we did not formulate rigorously the path-integration for such field theories. We simply assumed an appropriate analytic continuation of the couplings of the relevant hermitian interactions 
 in order to describe the non-hermitian ones 
 in the path integral, formally, without an attempt to define properly the corresponding fermion and (pseudo)scalar path-integral measures and discuss renormalization group properties. 
Such a programme has been initiated in \cite{bs} for the simple case of a PT symmetric Hamiltonian of a pseudoscalar field. Extension of this approach to our field theoretic system, which involves both fermions and (pseudo)scalar fields is still pending, and in our case is complicated by the fact that in the case of physical interest, we have a Yukawa interaction between pseudoscalar fields (axions) and fermions.  It will be interesting to see, by extending the approach of \cite{bs} to our systems, whether the appropriate renormalization group considerations of the non-hermitian interactions in our models will affect significantly the SD results for dynamical generation presented here and in \cite{amsot}. 

Moreover, it would also be of great interest to understand the microscopic nature of the non-hermitian anomaly terms themselves ({\it cf.} \eqref{bacoupl3}), for the case of massless chiral fermions,  independently of their role on mass generation.
In hermitian theories,  it is well known that the anomalies carry topological information, being associated 
with the index of the pertinent Dirac operator for fermions in curved space times~\cite{anomalies}. A natural question concerns the precise formulation  of non-hermitian Hamiltonians of, say, sterile neutrinos in such curved space times in the presence of gravitational anomalies (and, of course, more generally, of Dirac fermions in the presence of non-hermitian mixed -gravitational and gauge- anomalies), and whether such topological properties persist in such cases. 

We hope to tackle such issues in future publications.

\section*{Acknowledgements}

We thank Jean Alexandre for discussions. 
AS wishes to thank the Department of Physics of King's College London for a visiting doctoral student appointment, during which the current work was completed. 
The work  of NEM is supported in part by the UK Science and Technology Facilities  research Council (STFC) under the research grants ST/P000258/1 and 
ST/T000759/1. The work of AS is supported by the CONICYT-PFCHA/Doctorado Nacional/2017-21171194. NEM also acknowledges a scientific associateship (``\emph{Doctor Vinculado}'') at IFIC-CSIC-Valencia University, Valencia, Spain.

\end{document}